\documentclass[fleqn,usenatbib]{mnras}

\usepackage{newtxtext,newtxmath}
\usepackage[T1]{fontenc}
\usepackage{rotating}
\usepackage{pdflscape}

\DeclareRobustCommand{\VAN}[3]{#2}
\let\VANthebibliography\thebibliography
\def\thebibliography{\DeclareRobustCommand{\VAN}[3]{##3}\VANthebibliography}
\usepackage{graphicx}	
\usepackage{amsmath}	

\title{A giant dark matter core extending far beyond the stars of a dwarf galaxy}

\author[Q. Yuan et al.]{Qinwei Yuan$^{1}$, Yong Shi$^{2,*}$, Xin Li$^{2}$, Zhiyu Zhang$^{1}$, Zhiyuan Zheng$^{3}$, Qiusheng Gu$^{1}$, Xiaoling Yu$^{4}$,
\newauthor Fuyan Bian$^{5}$ \\
$^{1}$School of Astronomy and Space Science, Nanjing University, Nanjing 210093, China\\
$^{2}$Department of Astronomy, School of Science, Westlake University, Hangzhou, Zhejiang 310030, China\\
$^{3}$Research Center of Astronomy, Qinghai University, Xining, 810016, China\\
$^{4}$College of Physics and Electronic Engineering, Qujing Normal University, Qujing, 655011, China\\
$^{5}$European Southern Observatory, Alonso de C\'ordova 3107, Casilla 19001, Vitacura, Santiago 19, Chile\\
$^{*}$Corresponding author(s). E-mail(s): shiyong@westlake.edu.cn
}

\date{Accepted XXX. Received YYY; in original form ZZZ}

\pubyear{\the\year{}}

\begin{document}
\label{firstpage}
\pagerange{\pageref{firstpage}--\pageref{lastpage}}
\maketitle

\begin{abstract}
In the standard cosmological model, the distribution of cold dark matter (CDM) can be reshaped by baryonic processes through gravity, as gas outflows alter the central potential and transform steep cusps into nearly constant-density cores. However, such feedback-generated cores are generally expected to have sizes comparable to the visible stellar body of a galaxy. Here we report the discovery of an exceptionally extended dark matter core in a nearby gas-rich dwarf galaxy UGCA 320. We combine high-resolution MUSE H$\alpha$ kinematics in the inner region with high-, medium-, and low-resolution MeerKAT HI observations from the MHONGOOSE survey tracing the galaxy out to large radii. After accounting for an additional non-circular kinematic component identified in the central HI spectra, we construct an extended rotation curve to constrain the dark matter distribution. The resulting rotation curve is poorly reproduced by a cuspy Navarro–Frenk–White (NFW) halo but well described by cored pseudo-isothermal (ISO) and Burkert halo models. Consistently, a fit with a generalized NFW model yields an inner density slope of $\gamma=0.36\pm0.12$, quantitatively supporting a shallow central density profile. The best-fit ISO model gives a dark matter core radius of $7.62\pm0.80$ kpc, approximately four times the stellar half-light radius and substantially larger than typically found in dwarf galaxies. We investigate the main systematic uncertainties and find that the resulting core radii remain large, all lying within 3$\sigma$ of the fiducial fit. The exceptionally large core in UGCA 320 may pose a challenge to the standard CDM paradigm. While its properties are broadly consistent with self-interacting dark matter (SIDM), they may be difficult to reconcile with warm and fuzzy dark matter models, as well as the stellar feedback-driven core formation scenario.
\end{abstract}

\begin{keywords}
Galaxies: dwarf -- galaxies: kinematics and dynamics -- dark matter
\end{keywords}

\medskip

\section{Introduction}

Since discoveries that galaxies have flat rotation curves at large radii (e.g., \citealt{Freeman1970, Rubin1970}) rather than follow Kepler's law, 
dark matter was proposed to account for the invisible mass (e.g., \citealt{Ostriker1974, Einasto1974}). In the standard $\Lambda$CDM cosmology, small-scale structures collapse first and then hierarchically merge into larger systems \citep{press1974, springel2006}. While this model successfully explains the large-scale structures of the Universe, it faces significant challenges on small scales \citep{Weinberg_2015, Bullock_2017, sales2022baryonicsolutionschallengescosmological}. 

These small-scale tensions emcompass the cusp-core problem \citep{Moore1994, Flores1994}, the diversity problem \citep{Oman_2015}, the missing satellites problem \citep{Klypin_1999, Moore_1999}, and the too-big-to-fail problem \citep{Boylan_Kolchin_2011}. Some of these issues, notably the missing satellites problem, have been alleviated by deeper observations and higher-resolution simulations that incorporate baryonic physics (e.g., \citealt{Willman_2005, Bechtol_2015, Laevens_2015a, Buck2019, santossantos2026}). Others, however, remain unresolved. In this paper, we focus on the cusp-core problem. In dissipationless cold dark matter N-body simulations, the central density profiles of dark matter halos follow a cuspy form, well described by the Navarro-Frenk-White (NFW) model \citep{Navarro1996,Navarro1997} as $\rho \propto r^{-1}$. In contrast, rotation curves derived from HI observations of dwarf and low surface brightness galaxies are shallower than these predictions and infer a constant-density core \citep{Moore1994,deblok1997,Oh_2011}, but with few exceptions \citep[e.g.][]{Shi_2021}. To mitigate potential beam smearing effects, high-resolution long-slit H$\alpha$ data \citep{swaters2000,McGaugh_2001,deBlok2001,spekkens2005,zackrisson2006} have been used to derive rotation curves, yielding results consistent with those from HI. Furthermore, subsequent studies employing high-resolution two-dimensional velocity fields \citep{de_Naray_2008,Oh_2015,Korsaga_2019} have eliminated concerns related to pointing offsets and non-circular motions, and they conclude that cored dark matter density profiles provide significantly better fits to the data than cuspy ones.

To address this long-standing tension encountered in $\Lambda$CDM scenario, various alternative dark matter candidates --- such as warm dark matter, self-interacting dark matter, and fuzzy dark matter --- have been proposed. Warm dark matter possesses a non-zero thermal velocity, which erases the primordial power spectrum on small scales via free-streaming effect and lowers the central dark matter density due to an upper limit on phase-space density \citep{Hogan_2000, Bode_2001, Macci__2012, Viel_2013}. For self-interacting dark matter, its particles can transfer energy through elastic scatterings, turning a cusp into a core; moreover, cored halos are less bound than their cuspy counterparts of the same mass, making substructures more susceptible to tidal stripping \citep{Spergel_2000, Kaplinghat_2016, Tulin_2018, Adhikari_2025}. With an ultralight mass, fuzzy dark matter exhibits prominent wave-like properties, and the quantum pressure predicted by quantum mechanics resists gravitational collapse, giving rise to a stable soliton core in the center \citep{Hu_2000, Schive_2014, eberhardt2025ultralightfuzzydarkmatter}. A detailed introduction to dark matter can be found in the Discussion.

As the discrepancy originates from dark-matter-only simulations that include only gravitational interactions, baryonic processes offer a possible solution within the $\Lambda$CDM framework. One well-known mechanism is the non-adiabatic gravitational potential fluctuations driven by multiple, repeated bursts of star formation, which heat the orbits of dark matter particles and flatten the central dark matter density\citep{Navarro_1996, Governato_2010, Pontzen_2012, Teyssier_2013, Collins_2022}. Subsequent cosmological simulations \citep{Di_Cintio_2013} have confirmed that the inner density slope sensitively depends on the stellar-to-halo mass ratio $M_*/M_{\mathrm{halo}}$, and such feedback-induced cores can be efficiently produced when $M_*/M_{\mathrm{halo}} \sim 5\times10^{-3}$ --- i.e., when the potential well is relatively shallow and the supernova feedback is strong enough to expel large amounts of gas (see also \citealt{Chan_2015, Tollet_2016, Macci__2020, Lazar_2020, Straight_2025}). Moreover, the sizes of such feedback-induced dark matter cores are generally found to be smaller than or comparable to the stellar distribution \citep{Read_2016,Lazar_2020}. Another heating mechanism involves the infall of compact, massive baryon clumps, which transfer energy and angular momentum to background dark matter particles in the central region via dynamical friction. This process can operate before stellar feedback becomes effective \citep{El_Zant_2001, El_Zant_2004, Nipoti_2014, Del_Popolo_2016}.

\begin{figure}
    \centering
    \makebox[\linewidth][c]{%
     \includegraphics[width=\linewidth]{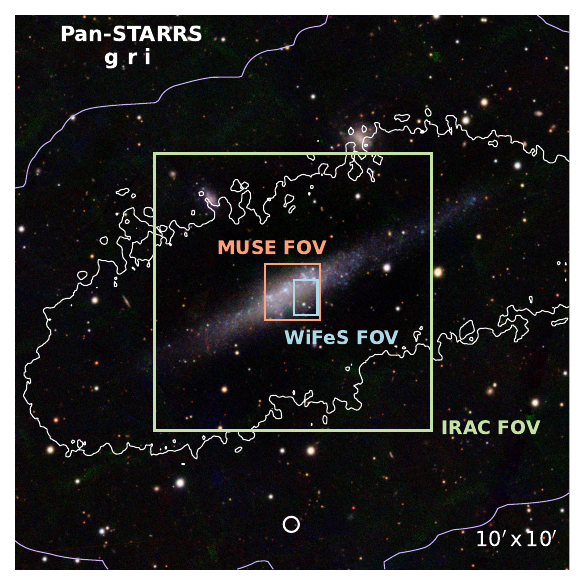}
    }
    \caption{False-color image of UGCA 320, constructed from Pan-STARRS g, r, and i band images, spanning an angular size of $\mathrm{10^{\prime} \times 10^{\prime}}$. The blue, red, and green rectangles mark the fields of view of WiFeS, MUSE, and IRAC, respectively. The white and purple contours show the HI intensity at a level of 2 $\mathrm{mJy\,beam^{-1}\, km\,s^{-1}}$ from the highest- and lowest-resolution moment-0 maps. The white ellipse at the bottom indicates the beam size ($8.2^{\prime\prime}\times7.1^{\prime\prime}$) of the highest-resolution HI data cube \texttt{r00\_t00}.}
    \label{falseimage}
\end{figure}

Dwarf galaxies serves as key laboratories for probing in small-scale challenges in the $\Lambda$CDM paradigm. On the one hand, they are dark-matter-dominated, offering an ideal testing ground for investigating dark matter distributions and constraining its nature. On the other hand, their shallow potential wells render them particularly susceptible to baryonic processes, making them excellent targets for studying the interplay between baryons and dark matter. This combination of these properties positions dwarf galaxies as optimal probes of the cusp-core problem, although it also introduces certain controversies and degeneracies \citep{Koudmani_2025}. In this work, we focus on a nearby gas-rich dwarf galaxy, UGCA 320, with a $M_{*}\,\sim\,10^{8.31}\,\mathrm{M_{\odot}}$ and a $M_{\mathrm{HI}}\,\sim\,10^{8.96}\,\mathrm{M_{\odot}}$ (Figure~\ref{falseimage}). To reduce systematic uncertainties, we leverage high-sensitivity HI observations, complemented by high-resolution H$\alpha$ data, and apply a three-dimensional modelling tool to robustly constrain the mass distribution. The basic properties of UGCA 320 are listed in Table~\ref{whole properties}. It has a redshift of about 0.0025 and a precise distance of $6.03^{+0.29}_{-0.21}$ Mpc measured via the Tip of the red giant branch (TRGB) \citep{Karachentsev2017}.

In this work, we assume a $\mathrm{\Lambda CDM}$ cosmology of $H_0=73\,\mathrm{km\,s^{-1}\,Mpc^{-1}}$, $\Omega_{0}=0.27$ and $\Omega_{\Lambda}=0.73$. This paper is structured as follows: Section 2 describes the observations of UGCA 320 and our data reduction procedures. The methods and results are presented in Section 3 and Section 4, respectively. The implications of our findings are discussed in Section 5, and the conslusions are summerized in section 6.

\begin{table}
\centering
\caption{The physical properties of UGCA 320}
\renewcommand{\arraystretch}{1}
\label{whole properties}
\begin{tabular}{c c c}
 \hline
 \hline
 Parameters & Value & Reference \\
 \hline
 R.A.(J2000) & 13:03:16.74 & \cite{Dale2009} \\
 Decl.(J200) & -17:25:22.9& \cite{Dale2009} \\
 Redshift & 0.0025 & \cite{li2025} \\
 Distance (Mpc) & 6.03 & \cite{Karachentsev2017} \\
 $\mathrm{log}(M_{*})\,[\mathrm{M_{\odot}}]$ & 8.31 & This work \\
 $\mathrm{log}(M_{\mathrm{HI}})\,[\mathrm{M_{\odot}}]$ & 8.96 & This work \\
 Inclination angle (\textdegree) & $74\pm4$ & This work \\
 Position angle (\textdegree) & $296\pm3$ & This work \\
 $V_{\mathrm{sys}}\,(\mathrm{km/s})$ & 739.9 & \cite{de_Blok_2024} \\
 \hline
\end{tabular}
\end{table}

\section{OBSERVATIONS AND DATA ANALYSIS}
 \subsection{HI Observations}
UGCA 320 is one of the targets in MHONGOOSE (MeerKAT HI Observations of Nearby Galactic Objects: Observing Southern Emitters) survey \citep{de_Blok_2024}, conducted with the MeerKAT radio telescope in South Africa. MeerKAT, a precursor to the Square Kilometer Array (SKA), consists of 64 antennas each 13.5 m in diameter \citep{deblok2016, Jonas:2018Jr}. With a minimum baseline of 29 m and a maximum baseline of up to 7.7 km, MeerKAT can simultaneously achieve high spatial resolution image and low column density sensitivity---a capability that previous HI surveys lacked \citep{de_Blok_2024}. The combination of MeerKAT's excellent column density sensitivity, high resolution, and large field of view (a primary beam FWHM of 1$^\circ$) enables MHONGOOSE to resolve cold accreting gas even in the outskirts of galaxies and study its effects on galaxy evolution \citep{deblok2016, de_Blok_2024, makinson2025}.  

The observations of MHONGOOSE survey began in October 2021. Every galaxy was observed for a total of 55 hours (full-depth), divided into ten tracks of 5.5 hours each (single-depth). The data were processed using the Containerized Automated Radio Astronomy Calibration (CARACal) pipeline \citep{caracal2020}. The survey produces full-depth data cubes with a velocity resolution of 1.4 $\mathrm{km\, s^{-1}}$ and angular resolutions ranging from $\mathrm{\sim\,8''}$ to $\sim\,90''$, using different weightings and taperings. The resulting column density sensitivity ($\mathrm{3\sigma}$ over 16 $\mathrm{km\, s^{-1}}$) varies from $\mathrm{\sim 10^{19}\,cm^{-2}}$ at the highest spatial resolution, to $\mathrm{\sim 10^{17}\,cm^{-2}}$ at the lowest. The flux densities measured in the MHONGOOSE survey are in good agreement with those from the single-dish Green Bank Telescope (GBT), with a flux ratio of GBT and MHONGOOSE of $1.058\pm0.080$, indicating that MHONGOOSE recovers the total flux within uncertainties. Further details are presented in \citet{de_Blok_2024}. 

\begin{figure*}
    \centering
    \includegraphics[width=\linewidth]{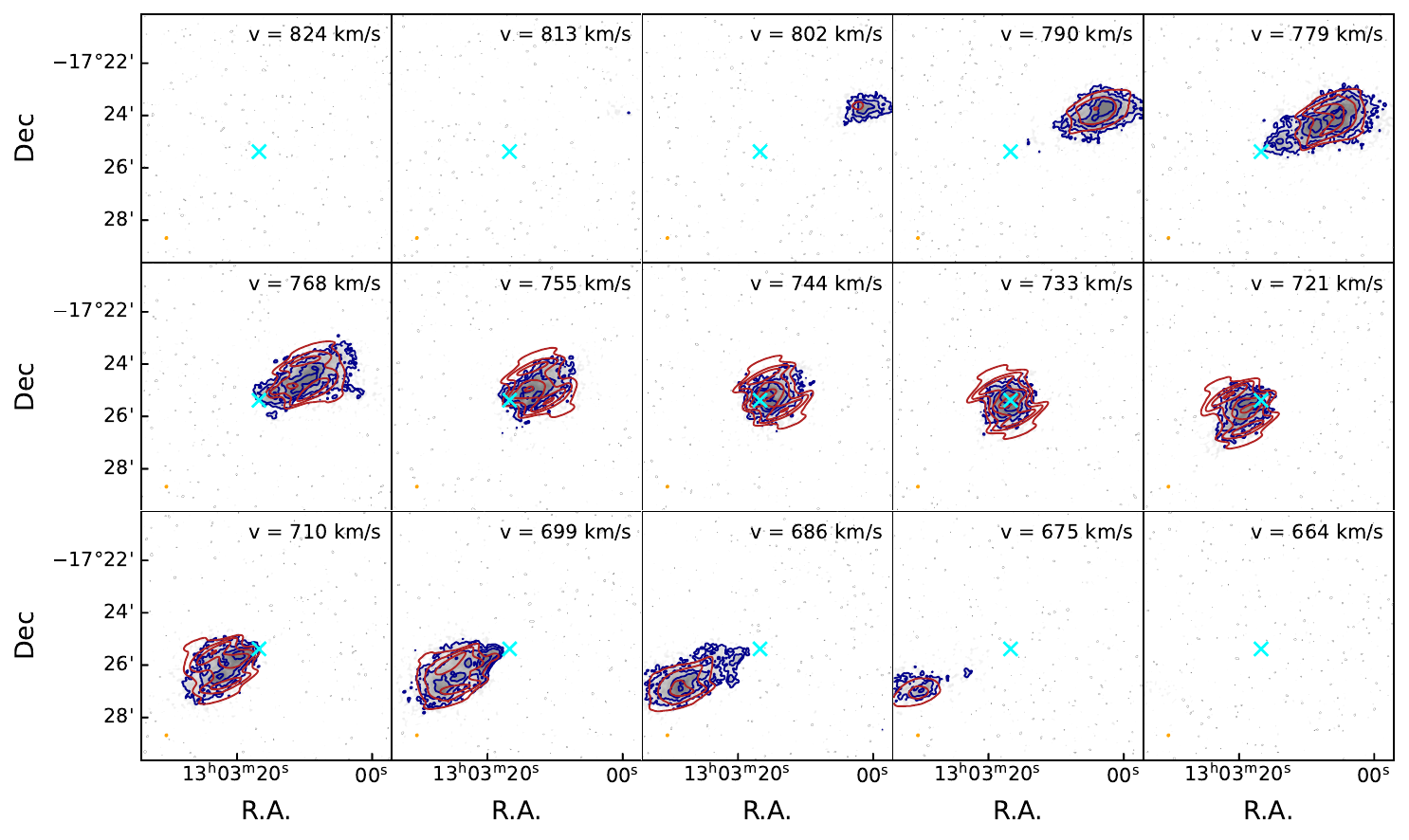}
    \caption{Channel maps of UGCA 320 for the highest-resolution \texttt{r00\_t00} cube, showing the observations (blue contours), and the $\rm{^{3D}BAROLO}$ model (red contours) after subtracting the outflow component. The cyan cross marks the galactic center, and the orange ellipse on the lower-left corner indicates the beam ($\mathrm{8.2^{\prime\prime}\times  7.1^{\prime\prime}}$). The velocity of the each channel is given in the upper-right corner. The contour levels are $2\sigma$, $4\sigma$, $8\sigma$, and $16\sigma$.}
    \label{chanmap}
\end{figure*}

In this paper, we use three data cubes for UGCA 320 (HIPASS J1303-17b): \texttt{r00\_t00}, \texttt{r05\_t60}, and \texttt{r10\_t90}. Their angular resolutions are $8.2^{\prime\prime}\times7.1^{\prime\prime}$ ($240\,\rm{pc}\times 208\,\rm{pc}$), $65.3^{\prime\prime}\times64.0^{\prime\prime}$ ($1.91\,\rm{kpc}\times1.87\,\rm{kpc}$), and $94.2^{\prime\prime}\times91.2^{\prime\prime}$ ($2.75\,\rm{kpc}\times2.67\,\rm{kpc}$), corresponding to $\mathrm{3\sigma}$ column density sensitivities of $\mathrm{5.89\times 10^{19}\,cm^{-2}}$, $\mathrm{9.54\times 10^{17}\,cm^{-2}}$ and $\mathrm{5.9\times 10^{17}\,cm^{-2}}$, respectively. In the labels, \texttt{r} denotes the robust weighting value and \texttt{t} denotes the taper. Figure~\ref{chanmap} shows the channel map of the highest-resolution cube \texttt{r00\_t00}, which extends to $\sim$ 9.4 kpc, while the lowest-resolution cube \texttt{r10\_t90} reaches $\sim$ 11.4 kpc. The different sensitivities allow us to characterize the complete distribution of HI gas and to derive a comprehensive rotation curve and matter distribution.Table~\ref{diff reso} lists the properties of the three cubes.

\begin{table}
\centering
\caption{The properties of MHONGOOSE data cubes}
\renewcommand{\arraystretch}{1.5}
\label{diff reso}
\begin{tabular}{c c c c c c}
 \hline
 \hline
  label & robust & taper & $b_{\mathrm{maj}}$ & $b_{\mathrm{min}}$ & $N_{\mathrm{HI}}^{3\sigma,16\,\mathrm{km\,s^{-1}}}$\\
   & value & & ($^{\prime\prime}$) & ($^{\prime\prime}$) & ($\mathrm{cm^{-2}}$)\\
  (1) & (2) & (3) & (4) & (5) & (6)\\
  \hline
  \texttt{r00\_t00} & 0.0 & 0 & 8.2 & 7.1 & $\mathrm{5.89\times 10^{19}}$ \\
  \texttt{r50\_t60} & 0.5 & 60 & 65.3 & 64.0 & $\mathrm{9.54\times 10^{17}}$ \\
  \texttt{r10\_t90} & 1.0 & 90 & 94.2 & 91.2 & $\mathrm{5.9\times 10^{17}}$ \\
 \hline
\end{tabular}
\end{table}

 \subsection{\texorpdfstring{H$\alpha$}{Halpha} Observations}
 \label{Ha observation}
The H$\alpha$ observation is part of the Dwarf Galaxy Integral-field Survey (DGIS; \citealt{li2025}). We carried out VLT/MUSE Integral Field Unit (IFU) observations in Wide Field Mode (WFM) \citep{bacon2010}, which provides a $1'\times 1'$ field of view (FOV) sampled at $0.2^{\prime\prime}$ per pixel. The spectral range covers from 4650$\mathrm{\AA}$ to 9300$\mathrm{\AA}$, with a mean spectral resolution of $R\sim3000$. The observation was performed on April 22 and 23, 2021, covering the central $1'\times 1'$ region of the galaxy (Figure~\ref{falseimage}). The total on-source time was 4480 seconds across eight exposures, which reduced to 3920 seconds after excluding one poor-quality frame. The full width at half maximum (FWHM) of Point-Spread Function (PSF) is $1.02''$.

According to the MUSE User Manual, the spectral resolution at 6500$\mathrm{\AA}$ is 2482, corresponding to an instrumental broadening of $\sim 44\,\mathrm{km\,s^{-1}}$. This value is much larger than the velocity dispersions of most H$\alpha$ regions in UGCA 320. Therefore, we also obtained Wide-Field Spectrograph (WiFeS) observation, mounted on the ANU 2.3-m telescope. WiFeS provides a higher spectral resolution of $R\sim7000$ over a wavelength coverage of 3290$\mathrm{\AA}$ to 9120$\mathrm{\AA}$, with a FOV of $25^{\prime\prime}\times38^{\prime\prime}$ (see \citealt{Dopita_2007} for details). The WiFeS observation was carried out on May 24, 2025, pointing toward the north-west offset from the galactic center along major axis (Figure~\ref{falseimage}), with an exposure time of 1200 seconds. The observing mode was nod and shuffle, using the B7000 and R7000 grisms with the dichroic RT560.

The raw data were reduced using the MUSE pipeline \citep{Weilbacher_2020} and PyWiFeS pipeline \citep{Childress_2013}, respectively. Both pipelines perform basic calibrations including wavelength and flux calibration, as well as corrections for bias, dark, flat, and skyflat frames. Cosmic ray removal, telluric correction, and atmospheric extinction correction are also included. For MUSE post-processing, we used Pan-STARRS broad-band images \citep{chambers2019panstarrs1surveys} to perform residual sky subtraction and final flux calibration, and calibrated the spectral shape for individual frames. The final MUSE datacube was rebinned by $2\times2$ pixels. For WiFeS, we used only the R7000 grism spectra, which cover H$\alpha$ line. Both data sets were corrected for telescope motions, world coordinate system (WCS), and the Milky Way extinction using the extinction curve of \cite{Cardelli1989} with $A_V/E(B-V)\approx3.1$ and $E(B-V)$ from \cite{Schlegel_1998}. The stellar continuum was subtracted using pPXF (Penalized PiXel-Fitting; \citealt{Cappellari_2016}). Subsequently, each emission line was fitted with a single Gaussian profile using \texttt{curve\_fit}. A detailed description of the fitting procedure is given in \citet{li2025}.

 \subsection{\texorpdfstring{IRAC 3.6 $\mathrm{\mu m}$ Observations}{IRAC 3.6 micron observations}}
The 3.6 $\mathrm{\mu m}$ image of UGCA 320 was obtained from the Spitzer Local Volume Legacy (LVL) survey \citep{Dale2009}. IRAC is a four-channel camera that obtains simultaneous broad-band images at 3.6, 4.5, 5.8, and 8.0 $\mathrm{\mu m}$. It has a large field of view of $5^{\prime}\times 5^{\prime}$ and a mean PSF FWHM of 1.6$^{\prime\prime}$ \citep{Fazio_2004}. UGCA 320 was observed on August 16, 2007 (Figure~\ref{falseimage}). Galactic extinction was corrected using the same method described in Section~\ref{Ha observation}. To mitigate flux contamination, we employed the \texttt{Python} package \texttt{photutils} to identify and subtract point sources following \cite{Dale2009}, specifically the blue foreground stars with flux ratios $f_{\nu}(3.6\,\mathrm{\mu m})/f_{\nu}(8\,\mathrm{\mu m})>8$, as well as extended background galaxies with smooth morphologies (ellipticity $>$ 0.8). After this, we manually masked some obvious bright point sources. In addition, we applied an extended source aperture correction following \cite{Dale2009}, because the IRAC flux calibration is based on point source photometry within a $12^{\prime\prime}$ radius aperture.

\section{METHODS}

 \subsection{The rotation curve of dark matter}
Baryons and dark matter both contribute to the total rotation curve, with the former comprising gas and stars:

\begin{equation}
    V_{\mathrm{tot}}^2=V_{\mathrm{gas}}^2+V_{\mathrm{star}}^2+V_{\mathrm{DM}}^2.
\end{equation}

\noindent In this subsection, we describe the methods used to extract the circular velocity from HI and H$\alpha$ data, and to determine the baryonic contributions.

  \subsubsection{The total rotation curve from HI data cubes}  
A common method to derive a rotation curve of a disc galaxy is the tilted-ring model. It assumes that the galactic disc can be described by a number of concentric rings, each with its own geometric and kinematic parameters, including kinematic center, inclination angle, position angle, systemic velocity and rotation velocity \citep{begeman1989}. We fitted the tilted-ring model to three HI data cubes with different angular resolution using $\rm{^{3D}BAROLO}$ \citep{diteodoro2015}, a software that fits 3D tilted-ring models to spectroscopic data cubes and accounts for beam smearing and asymmetric drift correction. 

\begin{figure}
    \centering
    \includegraphics[width=\linewidth]{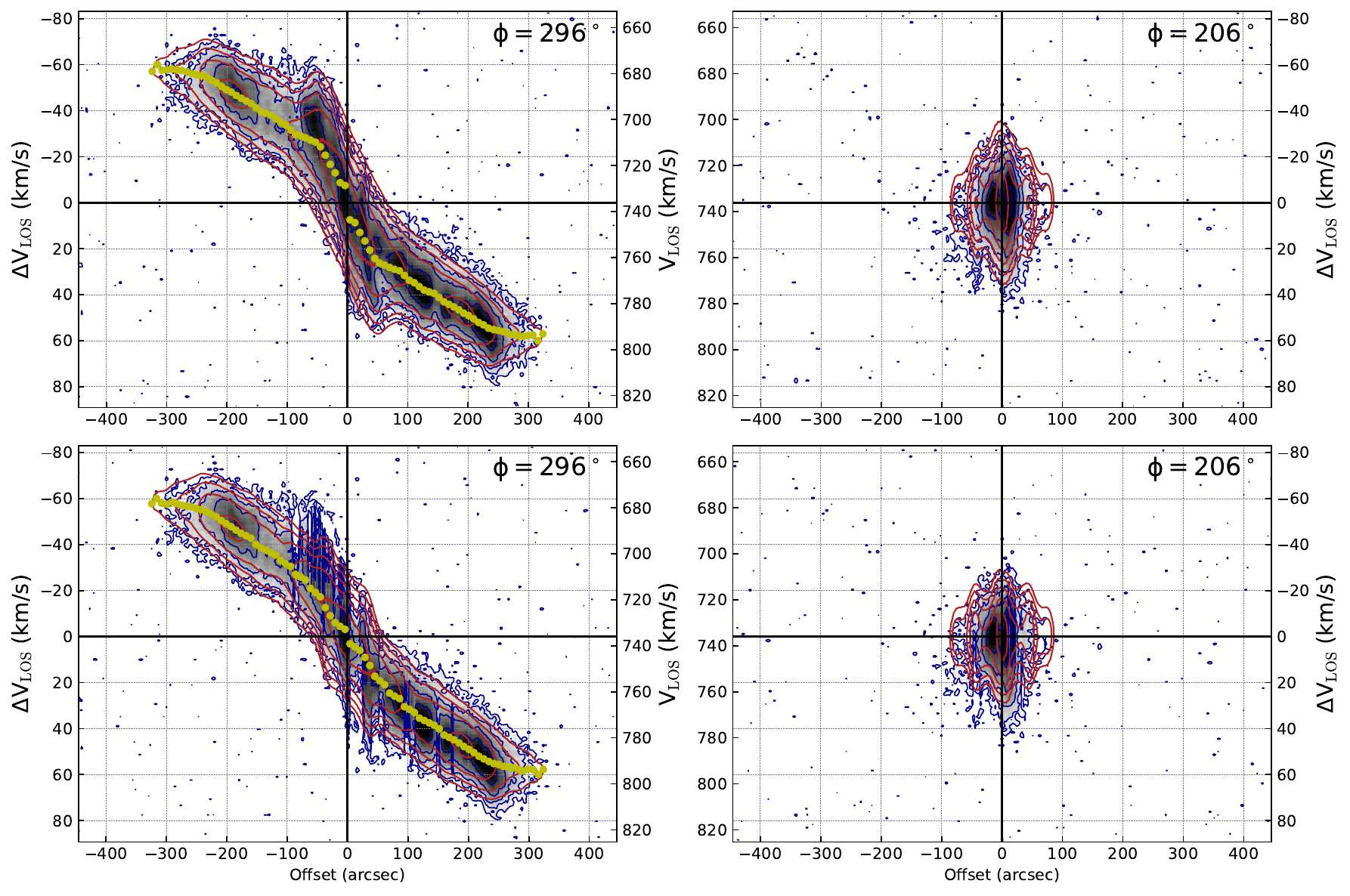}
    \caption{Position-velocity (PV) diagrams along major (left column) and minor (right column) axes of the original highest-resolution \texttt{r00\_t00} cube (upper row) and the decomposed cube after subtracting the outflow component (lower row). The overlaid blue and red contours represent the observed and the $\rm{^{3D}BAROLO}$ modelled HI intensity, respectively. The yellow circles indicate the line-of-sight velocity, and the position angle is given in the upper-right corner of each panel.}
    \label{pv}
\end{figure}

\begin{figure*}
    \centering
    \includegraphics[width=\linewidth]{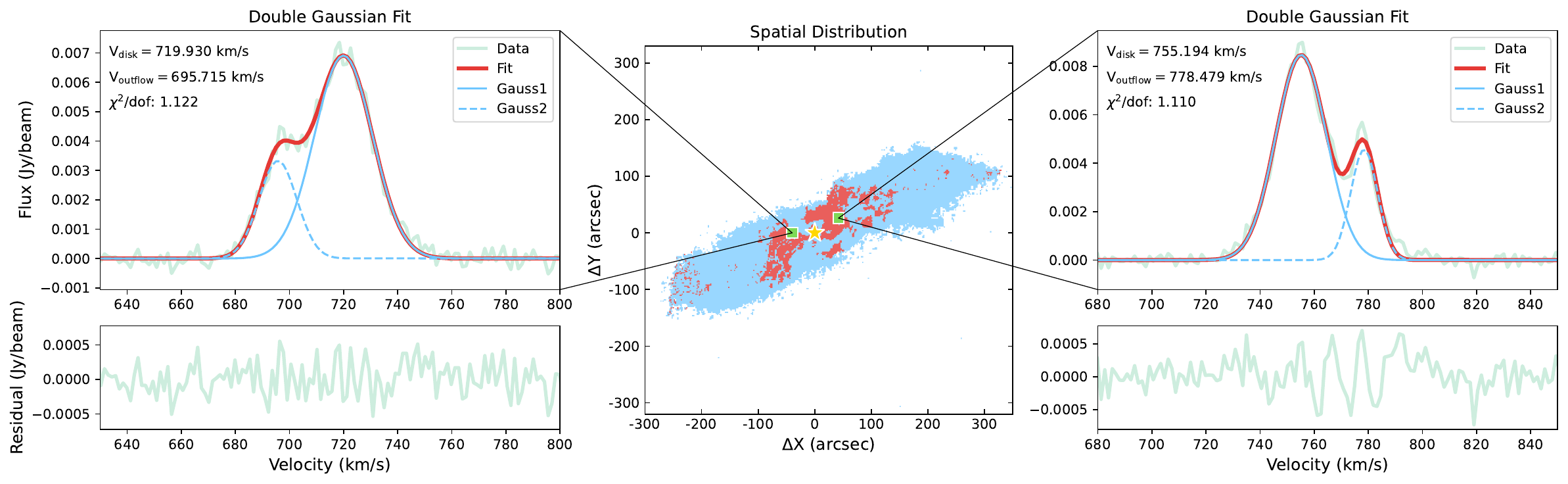}
    \caption{Examples of the double-peaked spectral profiles in the central region of UGCA 320. The middle panel displays the spatial distribution of the Gaussian fitting results: blue pixels represent regions with a single Gaussian component, while red pixels indicates regions requiring two Gaussian components. The gold star marks the galactic center, and the two green squares mark the positions of the example spectra. In the left and right panels, the upper and lower subpanels show the observed spectrum (light green) and the best-fit Gaussian profile (red), respectively. The two blue curves represent the two Gaussian components---one corresponds to the disk component and while the other to outflow.}
    \label{spec}
\end{figure*}

When running $\rm{^{3D}BAROLO}$ directly on the original highest-resolution data cube \texttt{r00\_t00}, we find that UGCA 320 exhibits complex kinematics, as also reported by \cite{alabi2025starsionizedgasugca} and \cite{zabel2026tracingneutralhydrogenugca}. From the position-velocity (PV) diagram along the major axis and the rotation curve produced by $\rm{^{3D}BAROLO}$, a prominent bump is visible in the central region (the upper panel of Figure~\ref{pv}), where the corresponding spectra show clear double-peaked profiles (Figure~\ref{spec}). \cite{zabel2026tracingneutralhydrogenugca} interpreted that this is plausibly an ongoing star-formation driven outflow.

To ensure that the HI spectra trace only the circular motion due to gravity, we performed spectral decomposition on the data cube \texttt{r00\_t00} to exclude additional component potentially driven by star formation. Since the MHONGOOSE survey does not provide a noise cube, we estimated the root mean square (RMS) by calculating the standard deviation of the line-free channels and assumed that the RMS to be constant across channels. We fitted both single- and double-Gaussian profiles to the spectra with a signal-to-noise (SNR) ratio greater than five in the velocity space, where the line-center velocity lies within $V_{\rm{sys}}\pm100\,\mathrm{km\,s^{-1}}$. The choice between single- and double-Gaussian models was based on the Bayesian Information Criterion (BIC) \citep{schwarz1978, Liddle_2007}:

\begin{equation}
    \mathrm{BIC} = -2\mathrm{ln}\mathcal{L}_{\mathrm{max}}+k\mathrm{ln}N,
    \label{BIC}
\end{equation}

\noindent where $\mathcal{L}_{\mathrm{max}}$ is the maximum likelihood of the model, $k$ is the number of parameters, and $N$ is the number of data points used in the fitting process. The first term of Eq.~(\ref{BIC}) represents the goodness of fit, while the second term penalize the number of parameters.

We adopted a double-Gaussian model when the following criteria were met: (1) $\Delta \mathrm{BIC}=\mathrm{BIC_{single}}-\mathrm{BIC_{double}}>10$; (2) the amplitudes of both Gaussian components have SNR larger than three; (3) the velocity separation between the two line centers exceeds the smaller of the two velocity dispersions, to ensure the two components are well resolved. We defined the component whose line center is farther from $V_{\mathrm{sys}}$ as the outflow component. The fitting results are shown in the middle panel of Figure~\ref{spec}, where red pixels indicate regions requiring two Gaussian components and blue pixels indicate regions with a single Gaussian component. The left and right panels present two examples of double-Gaussian fitting. We then subtracted the outflow component from the original data cube. For the \texttt{r05\_t60} and \texttt{r10\_t90} data cubes, the beam sizes are too large to resolve two components, therefore, we did not perform the same spectral decomposition for them.

For the \texttt{r00\_t00} data cube, we initialized the $\rm{^{3D}BAROLO}$ parameters using the photometric center from the Two Micron All Sky Survey (2MASS; \citealt{Skrutskie2006}) and position angle (\texttt{PA}) from elliptical isophote fitting of the IRAC 3.6 $\mathrm{\mu m}$ image. The kinematic center was fixed to this photometric center. To determine the inclination angle (\texttt{INC}), we explored several values from literature as well as the estimates from isophote fitting. By comparing the residuals between the data and the model given by $\rm{^{3D}BAROLO}$ across different runs, we set the initial \texttt{INC} to $74^{\circ}$. The initial systemic velocity (VSYS) was set to 739.9 $\rm{km\,s^{-1}}$, corresponding to the central velocity of the global spectrum reported by \citet{de_Blok_2024}. These parameters were left free during the fitting process.

As noted by \citet{makinson2025}, it is not physical to combine rotation curve derived at different resolutions while ignoring inconsistencies in the radial dependence of geometric parameters. For the kinematic center, fixing it for low-resolution data is reasonable, especially when the galaxy is highly inclined. In Figure~\ref{geometic}, we show the radial profiles of inclination and position angle at three different resolutions. The fluctuations in \texttt{INC} and \texttt{PA} are found to be about $\pm4^{\circ}$ and $\pm3^{\circ}$ in the outer disk, respectively. For consistency, we fixed the \texttt{CENTER}, \texttt{VSYS}, \texttt{INC}, and \texttt{PA} for \texttt{r05\_t60} and \texttt{r10\_t90} data cubes to the median values of the outputs from \texttt{r00\_t00}, as these parameters are less well constrained at lower resolutions. The associated systematic uncertainties will be further discussed.

For all three runs, the initial parameters for rotation velocity (\texttt{VROT}) and velocity dispersion (\texttt{VDISP}) were estimated as the mean values derived from the moment-1 and moment-2 maps. The scale height of the gas disc, \texttt{Z0}, was set to 100 pc following \citet{iorio2016} and \citet{Shi_2021}. For the \texttt{r00\_t00} cube, the radial separation between each rings (\texttt{RADSEP}) was set to $8.2^{\prime\prime}$, which corresponds to the FWHM of the beam major axis. For the \texttt{r05\_t60} and \texttt{r10\_t90} cubes, RADSEP was set to $45^{\prime\prime}$ and $60^{\prime\prime}$ respectively to obtain a sufficient number of data points. We limited the number of rings (\texttt{NRADII}) to prevent $\rm{^{3D}BAROLO}$ from extending into noise-dominated regions.

We chose \texttt{SIDE=B}, meaning that both sides of the galaxy are fitted. The normalization type was set to \texttt{AZIM}. We adopted two-stage fitting approach, in which a second fitting stage is performed after the inclination angle and position angle are regularized using a Bezier function. Other optional parameters were left as their default values. The specific values are listed in the Table~\ref{initpara}. 

\begin{figure*}
    \centering
    \includegraphics[width=\linewidth]{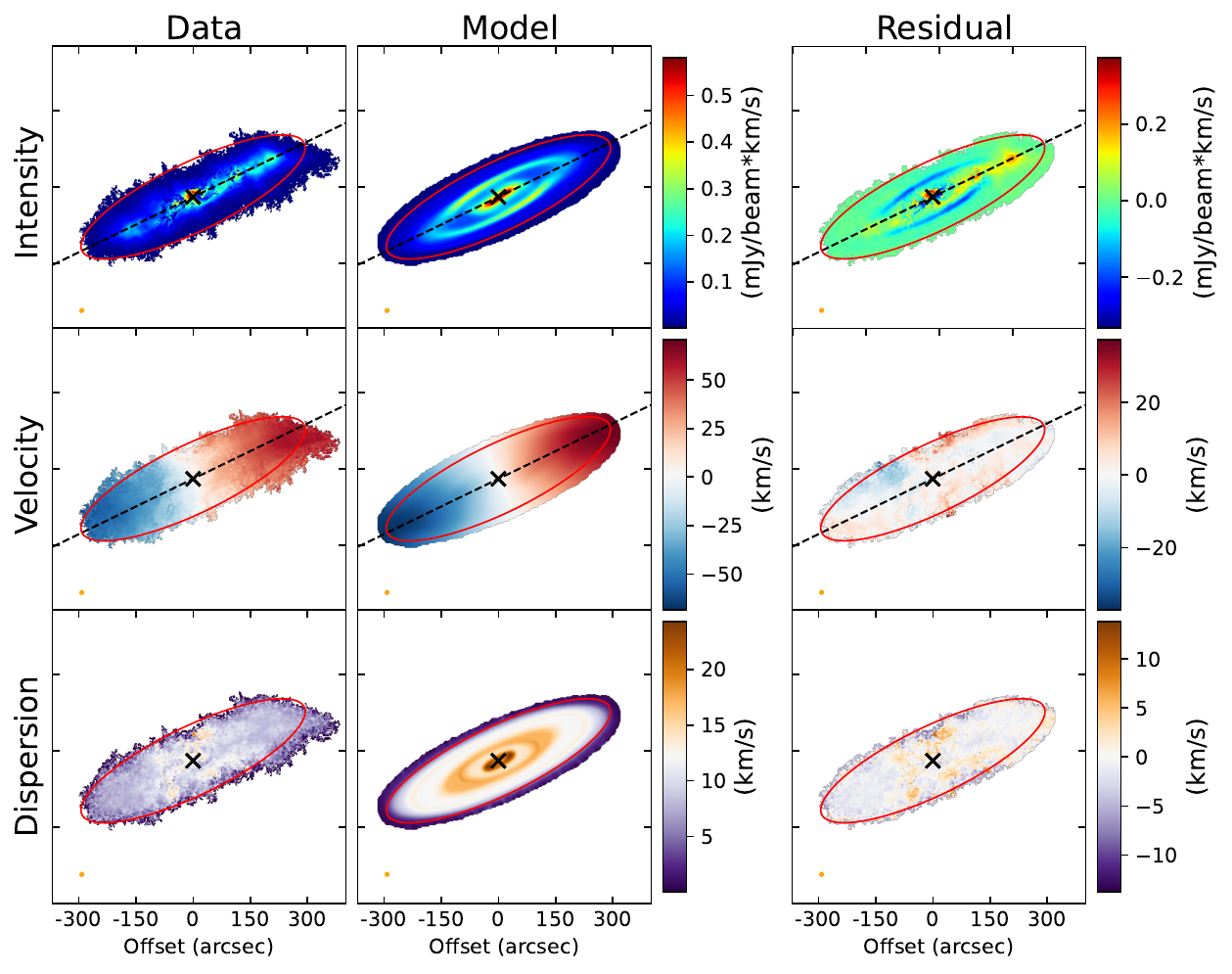}
    \caption{Moment maps of the highest-resolution HI data cube \texttt{r00\_t00} after decomposition, comparing the observations with the $\rm{^{3D}BAROLO}$ models. The left column shows the observed data after decomposition, the middle column the best-fit model, and right column the residual. From top to bottom, the three rows present the HI intensity, the line-of-sight velocity, and the velocity dispersion. The black cross marks the galactic center, while the orange ellipse in the lower-left corner indicates the beam ($\mathrm{8.2^{\prime\prime}\times  7.1^{\prime\prime}}$). The overlaid red ellipse traces the outermost ring used in the fitting.}
    \label{3DB_high_mom}
\end{figure*}

\begin{figure}
    \centering
    \includegraphics[width=\linewidth]{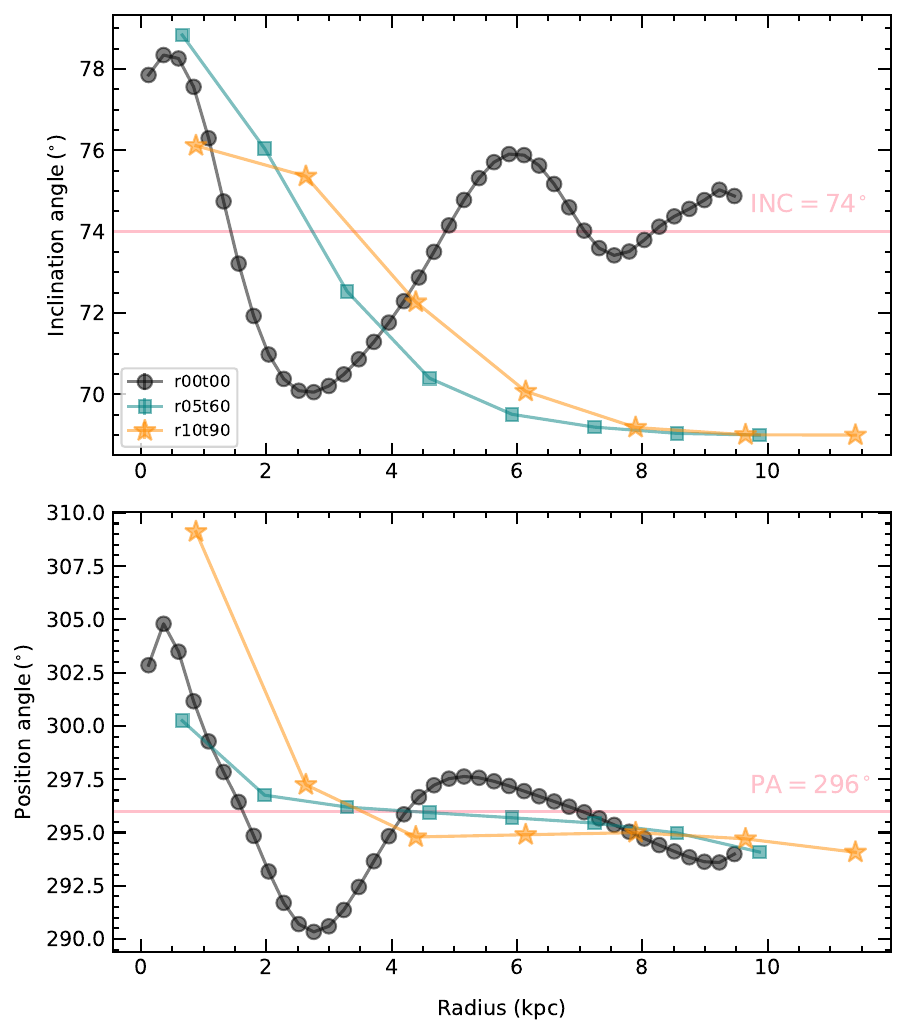}
    \caption{Radial profiles of the inclination angle and position angle at different resolutions, as derived from $\rm{^{3D}BAROLO}$ without fixing these parameters. Black circles, green diamonds, and gold stars represent the data points from the \texttt{r00\_t00}, \texttt{r05\_t60}, and \texttt{r10\_t90} cubes after regularization (i.e., fitting a B\'ezier function to the first stage results). The pink lines indicate the median values of \texttt{INC} and \texttt{PA} obtained from the \texttt{r00\_t00} fit.}
    \label{geometic}
\end{figure}

Figure~\ref{pv} compares the PV diagrams of the \texttt{r00\_t00} cube before and after decomposition. In the original diagram along major axis (upper right panel), the contours exhibit a symmetric bump in the central region, and the corresponding rotation curve appears warped, indicating the presence of non-circular motion. After subtracting the outflow (lower right panel), the contours become more regular and the rotation curve smoother, suggesting that our decomposition effectively recovers the pure circular motion governed by gravitational potential. Figure~\ref{3DB_high_mom} displays the moment maps derived from observations and tilted-ring modeling for \texttt{r00\_t00} cube. The median absolute velocity residual is $\sim2.8$ $\rm{km\,s^{-1}}$, while the outskirts along minor axis show the largest velocity residuals, up to $\sim37$ $\rm{km\,s^{-1}}$, consistent with the excess contours observed in the PV diagram along minor axis. Concerned that we might not subtract the outflow sufficiently, we tried adding radial velocity as a free parameter in the fitting process. However, the derived maximum radial velocity is only $\sim6$ $\rm{km\,s^{-1}}$, which still can not account for the remaining velocity residuals. This finding underscores the complex kinematics of UGCA 320. The PV diagrams and moment maps of \texttt{r05\_t60} and \texttt{r10\_t90} are shown in figure~\ref{pv_med_low} and figure~\ref{3DB_med_mom&low}. Because no spectral decomposition was performed for these two lower-resolution cubes, they exhibit relatively larger velocity dispersions in the galactic center. Overall, the models provide a good fit to the data, with small residuals.

For the $\texttt{r10\_t90}$ run, $\rm{^{3D}BAROLO}$ did not provide an uncertainty estimate for the rotation velocity, so we adopted 10\% of the rotation velocity as its uncertainty. The rotation curves derived from the three HI data cubes are shown in Figure~\ref{modelfit}. Due to differences in HI sensitivity, the rotation curves of \texttt{r00\_t00}, \texttt{r05\_t60}, and \texttt{r10\_t90} extend over different radial ranges. Since the lower resolutions of \texttt{r05\_t60} and \texttt{r10\_t90} cubes prevent us from performing the same decomposition as applied to \texttt{r00\_t00}, the inner data points from these two cubes are inconsistent with those derived from the \texttt{r00\_t00} cube. We therefore decided to use only the outer data points from \texttt{r05\_t60} and \texttt{r10\_t90} when fitting the dark matter models.

  \subsubsection{\texorpdfstring{The rotation curve from H$\alpha$ map}{The observed rotation curve from Ha map}}
H$\alpha$ emission, which traces ionized gas from recent star formation, has a spatial extent significantly smaller than that of HI. The central velocity gradient may appear flattened in HI observations due to beam smearing \citep{Noordermeer_2007, kam2017hikinematicsmassdistribution}. In contrast, with high resolution, H$\alpha$ provides stronger constraints on the rotation curve and the resulting dark matter density profile in the inner region, where the data are well sampled. Therefore, it is ideal to combine the H$\alpha$ and HI rotation curves to probe both the inner and outer regions more accurately \citep{Kam_2015, Korsaga_2019}. Unlike neutral hydrogen, H$\alpha$ emission is clumpy and concentrated around star-forming regions. Hence, rather than fitting a 3D tilted-ring model, we extracted the H$\alpha$ rotation curve along the major axis as \cite{begeman1989}:

\begin{equation}
    V_{\mathrm{obs}}(R)=V_{\mathrm{sys}}+V_{\mathrm{rot}}(R)\,\mathrm{cos\,}  \phi\sin i,
\label{vrot}
\end{equation}

\noindent where $\phi$ and $i$ refer to position angle and inclination angle, respectively.

First, we adopted an SNR threshold of five for the H$\alpha$ flux and used the dendrogram algorithm \citep{Rosolowsky2008} as implemented in the \texttt{astrodendro} \texttt{Python} package \citep{Robitaille2019} to identify H$\alpha$ clumps with sizes larger than the FWHM of PSF. The \texttt{min\_delta} parameter was set to the median value of the H$\alpha$ background to distinguish between neighboring clumps (Zheng in prep). Next, we applied an SNR threshold of seven and masked all candidate HII regions. The kinematic parameters derived from $\rm{^{3D}BAROLO}$ were employed to determine the major axis on the H$\alpha$ map, and the five pixels closest to the major axis were defined as the major-axis pixels. To improve the SNR, we rebinned the line-of-sight velocity using a bin size of eight times the pixel scale, corresponding to 3.2$^{\prime\prime}$, discarding bins with insufficient data points. The mean velocity and standard deviation within each bin were taken as the rebinned line-of-sight velocity and its associated error, respectively. We calculated the velocities on both the approaching and receding sides and then took their average.

In the region where velocity dispersion is high, random gas motion reduces the observed rotation velocity \citep{bureau2002, Oh_2015, iorio2016}. Therefore, it is crucial to apply an asymmetric drift correction to the rotation velocity in order to recover the circular velocity, which traces the gravitational potential. For a gas disk is in equilibrium, the following relation holds:

\begin{equation}
    V_{\mathrm{circ}}^2=V_{\mathrm{rot}}^2+V_{\mathrm{A}}^2,
\end{equation}

\noindent where $V_{\mathrm{rot}}$ and $V_{\mathrm{circ}}$ are the observed rotation velocity and the circular velocity corrected for the effect of pressure support, respectively, and $V_{\mathrm{A}}$ denotes the asymmetric drift correction. The correction can be calculated as:

\begin{equation}
\begin{split}
    V_{\mathrm{A}}^2 &=-R\sigma_v^2\frac{
    \mathrm{\partial ln}(\rho\sigma_v^2)}{\partial R} \\
    &= -R\sigma_v^2\frac{\partial \mathrm{ln}(\sigma_v^2\Sigma_{\mathrm{int}} z_d^{-1})}{\partial R},
\end{split}
\end{equation}

\noindent where $\sigma_v$ is the velocity dispersion, $\rho$ is the gas volume density, $\Sigma_{\mathrm{int}}$ is the intrinsic gas mass surface density, and $z_d$ is the vertical scale height. The second line follows by assuming that, in the equatorial plane, $\rho$ is proportional to $\Sigma_{\mathrm{int}}/z_d$ \citep{iorio2016}.

Assuming that the scale height of the gas disk is independent of radius and that the disk is ultra-thin, we can derive the following:

\begin{equation}
    V_{\mathrm{A}}^2 =-R\sigma_v^2\frac{\partial \mathrm{ln}(\sigma_v^2\Sigma_{\mathrm{obs}}{\mathrm{cos}}i)}{\partial R}.
\label{asym}
\end{equation}

Asymmetric drift correction is already implemented in $\rm{^{3D}BAROLO}$, so we do not need to apply it manually to the HI data. For H$\alpha$, however, we must perform the correction ourselves. We derive the H$\alpha$ velocity dispersion $\sigma_{\mathrm{H\alpha}}$ around the galactic center from the emission line fitting results of WiFeS datacube, and subtract a thermal broadening of 10 $\rm{km\,s^{-1}}$. Because the WiFeS FOV is limited, we estimate $\sigma_{\mathrm{H\alpha}}$ in the remaining regions by scaling the HI velocity dispersion $\sigma_{\mathrm{HI}}$ based on their ratio measured in the overlapping region. We find $\sigma_{\mathrm{H\alpha}}/\sigma_{\mathrm{HI}}\approx1.7$ along major axis within the central region. Additionally, we assume that the radial profile of H$\alpha$ mass surface density follows a similar trend as that of HI.

When calculating the radial derivative in Equation~\ref{asym}, we followed \citet{bureau2002} and \citet{iorio2016} by fitting $\sigma_v^2\Sigma_{\mathrm{obs}}\mathrm{cos}i$ with an analytical function:

\begin{equation}
    f(R)=f_0\mathrm{(\frac{R_c}{arcsec}+1)(\frac{R_c}{arcsec}+e^{\frac{R}{R_d}})^{-1}}
\end{equation}

\noindent to avoid the large fluctuations in the derivative caused by variations in the estimated velocity dispersion.

In figure~\ref{modelfit}, the red squares in the inner region show the rotation velocity curve derived from H$\alpha$. The H$\alpha$ and HI rotation curves are generally in good agreement over their common radial range.

  \subsubsection{The rotation curve of gas}
We divided the gas disk into a series of concentric rings sharing the same inclination angle, position angle, and axis ratio, while their major and minor axes increase progressively with radius. The values of galactic center, inclination angle, and position angle were taken from the $\rm{^{3D}BAROLO}$ results. We first measured the mean HI integrated intensity from the three different resolution HI moment-0 maps (intensity maps) using elliptical aperture photometry. The HI mass was then calculated following the formula from \citet{Walter_2008}:

\begin{equation}
    M_{\mathrm{HI}}=2.36\times10^5 D^2 \times \sum_i S_i \Delta v,
\end{equation}

\noindent where $M_{\mathrm{HI}}$ is in $\mathrm{M_{\odot}}$, and $\sum_i S_i \Delta v$ is the integrated intensity in $\mathrm{Jy\,km\,s^{-1}}$. $D$ denotes the distance in $\mathrm{Mpc}$. The HI mass surface density was derived by dividing the mass within each elliptical annulus by its corresponding area in $\mathrm{pc^2}$. To correct for projection effects, we applied a factor of $\mathrm{cos}i$. We then converted the HI mass surface density to the total atomic gas surface density by multiplying a factor of 1.36 to account for the helium contribution \citep{Leroy_2008}. The uncertainty in the mass surface density was computed via error propagation, incorporating both the background noise and the $10\%$ flux calibration uncertainty of MeerKAT \citep{Serra_2023}. The 1-$\sigma$ background noise of moment-0 map was determined pixel-by-pixel as $\Delta v \cdot \sigma_{\mathrm{ch}} \cdot \sqrt{N_{\mathrm{ch}}}$, where $\sigma_{\mathrm{ch}}$ is the channel noise measured from line-free channels and $N_{\mathrm{ch}}$ is the number of channels integrated at that position. The uncertainty in the inclination angle will be treated separately as a systematic uncertainty in Section 4.2.

\begin{figure}
    \centering
    \includegraphics[width=\linewidth]{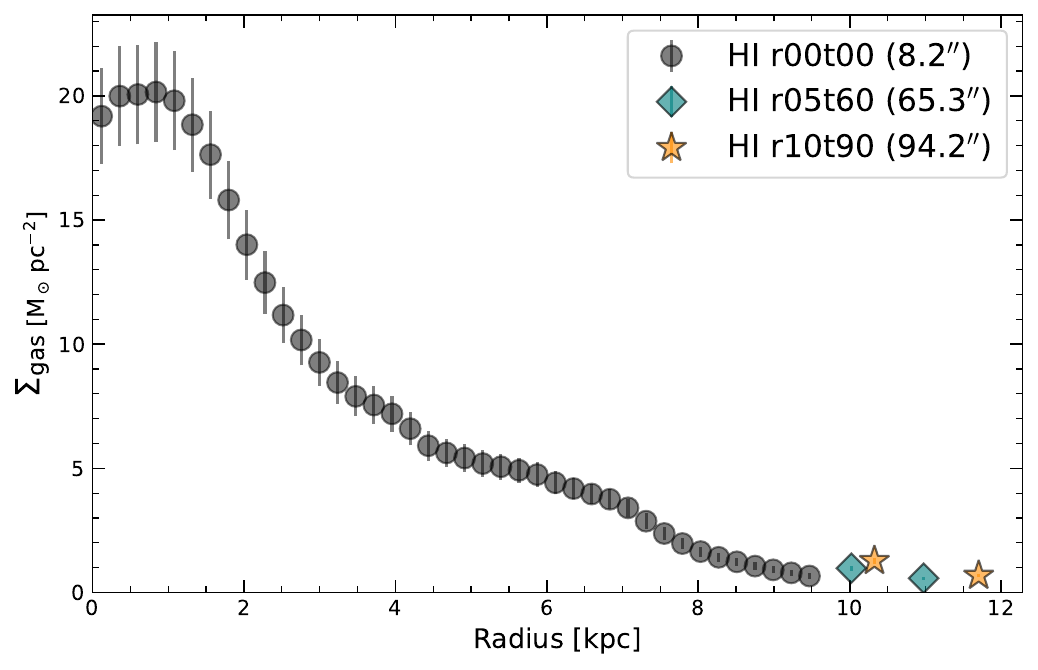}
    \caption{Multi-resolution radial profile of atomic gas mass surface density of UGCA 320. Data points from the high-, medium-, and low-resolution HI moment-0 maps are denoted by black circles, green diamonds, and yellow stars, respectively.}
    \label{gasmassdens}
\end{figure}

To construct a composite, multi-resolution radial profile of gas mass surface density, we combined data from the high-resolution moment-0 map (tracing inner region) with those from the lower-resolution maps (covering the larger radii). This composite profile, shown in Figure~\ref{gasmassdens}, was then used as the input for the Groningen Image Processing System (\texttt{GIPSY}) \citep{vanderHulst1992} task \texttt{ROTMOD} to derive the gas rotation curve. In the \texttt{ROTMOD} modeling, we set the disk vertical density distribution (\texttt{ZLAW}) to \texttt{SECH-SQUARE} and fixed the gas disk scale height (\texttt{Z0}) to 0.1 kpc. To account for uncertainties, we generated 1000 realizations of the gas mass surface density by perturbing the densities within their errors. Each realization was then used to compute a corresponding rotation curve with \texttt{ROTMOD}. Finally, the median value of the 1000 curves at each radius was adopted as the gas rotation velocity, while the standard deviation was taken as its uncertainty. The resulting gas rotation curve is shown in Figure~\ref{modelfit} as green lines.

  \subsubsection{The rotation curve of stars}
Although old stars are individually less massive, their overwhelmingly larger numbers relative to massive young stars make them the dominant contributors to the total stellar mass in galaxies. In addition, near-infrared (NIR) images are less affected by dust extinction than optical ones, and the stellar mass-to-light ratio is nearly constant in the NIR band \citep{McGaugh2014}. Therefore, we employed the IRAC 3.6 $\mathrm{\mu m}$ image to derive the stellar mass distribution of UGCA 320. Elliptical aperture photometry was performed, excluding masked pixels, to extract the radial profile of the 3.6 $\mathrm{\mu m}$ surface brightness. Because contamination from PAH 3.3 $\mathrm{\mu m}$ is negligible in low-metalicity environments \citep{Madden2006}, we did not consider it. For the uncertainties, we took into account the background noise provided by LVL and a 10$\%$ IRAC calibration uncertainty \citep{Reach_2005}. 

Given the large uncertainties for data points beyond 2 kpc, we chose to fit the radial intensity profile with a S\'ersic function and used the fitted values for the outer region. The best-fit half-light radius $R_\mathrm{e}$ and S\'ersic index $n$ are $2.02\pm0.10$ kpc and $0.62\pm0.06$, respectively, suggesting that UGCA 320 lacks a significant bulge component.

To convert the IRAC flux from $\mathrm{MJy\,sr^{-1}}$ to surface brightness in $\mathrm{mag\,arcsec^{-2}}$, we applied the formula from \citet{Oh2008}:

\begin{equation}
    \mu_{3.6\mu m}=-2.5\times \mathrm{log}_{10}(\frac{S_{3.6\mu m}\times 2.35 \times 10^{-5}}{280.9}),
\end{equation}

\noindent where $S_{3.6\mu m}$ is the 3.6 $\mathrm{\mu m}$ flux density in $\mathrm{MJy\,sr^{-1}}$, and 280.9 is the IRAC zero magnitude flux density in Jy \citep{Reach_2005}. Next, we transformed the surface brightness from $\mathrm{mag\,arcsec^{-2}}$ into mass surface density in $\mathrm{M_{\odot}\,pc^{-2}}$ following \citep{Oh2008}:

\begin{equation}
    \mathrm{\Sigma_{star} = \gamma_{3.6}\times10^{-0.4\times(\mu_{3.6\mu m}-24.8)}\times cos}i,
\end{equation}

\noindent where 24.80 is a conversion factor from $\mathrm{mag\,arcsec^{-2}}$ to $\mathrm{L_{\odot}\,pc^{-2}}$.

\begin{figure}
    \centering
    \includegraphics[width=\linewidth]{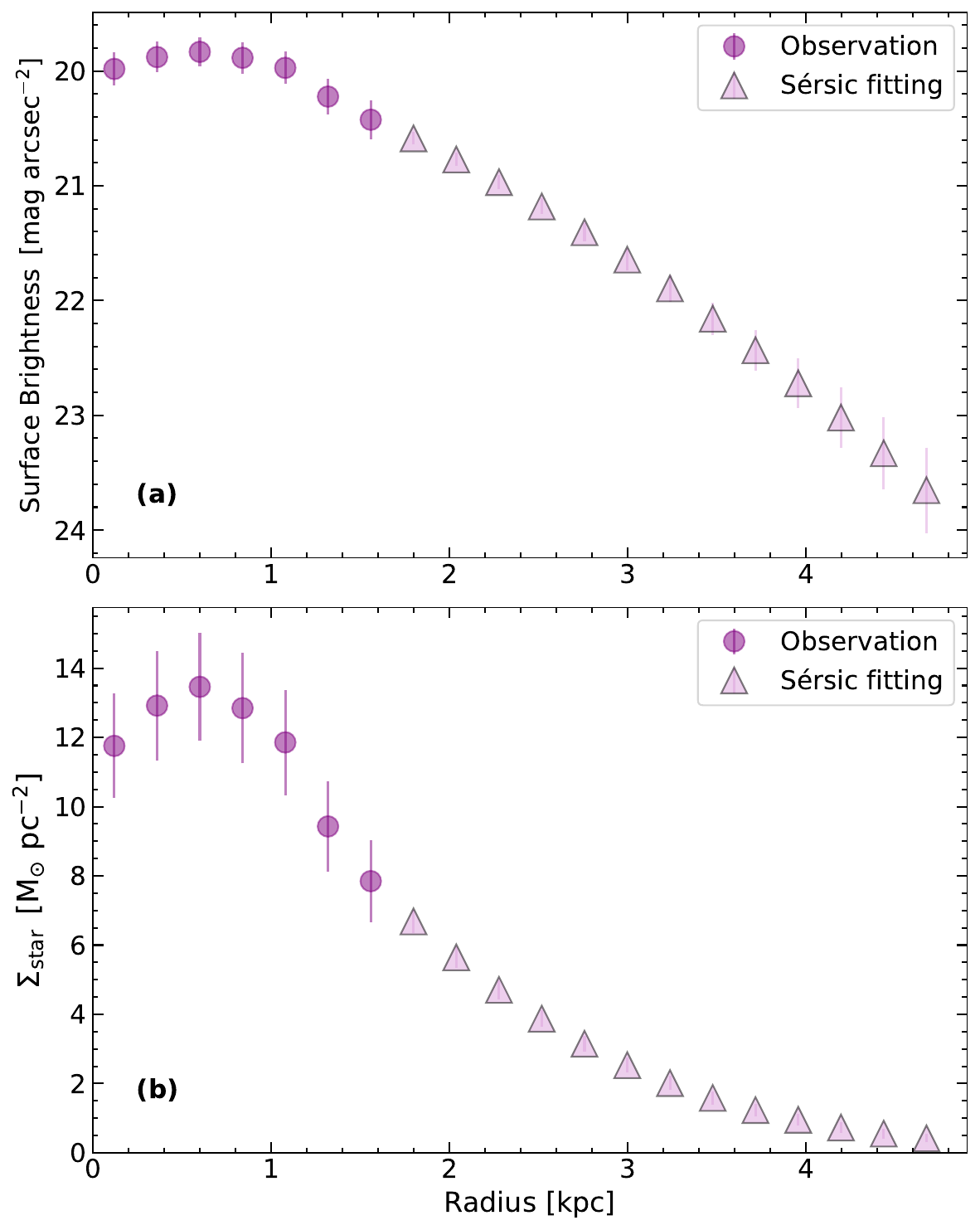}
    \caption{Radial profiles of the 3.6 $\rm{\mu m}$ surface brightness (a) and the stellar mass surface density (b) of UGCA 320. The dark purple circles represent the observed data points, while the light purple triangles denote the values from S\'ersic fitting.}
    \label{starmassdens}
\end{figure}

We then turned this surface brightness into the stellar mass surface density by adopting a stellar mass-to-light ratio $\gamma_{3.6}=0.5\,\mathrm{M_{\odot}/L_{\odot,3.6\mu m}}$ \citep{Cook2014, Lelli_2016}, where $\mathrm{L_{\odot,3.6\mu m}=1.4\times 10^{32}\,erg\,s^{-1}}$ \citep{Oh2008} is the solar luminosity at 3.6 $\mathrm{\mu m}$. The stellar mass of UGCA 320 is estimated to be $10^{8.31}\,\mathrm{M_{\odot}}$. The radial profiles of surface brightness and stellar mass surface density are displayed in Figures~\ref{starmassdens} (a) and (b), respectively.

We applied the same method to derive the stellar rotation curve. For the \texttt{ROTMOD} runs, \texttt{ZLAW} was set to \texttt{SECH-SQUARE} and \texttt{Z0} was fixed to 0.2 kpc. The uncertainties in the rotation curve were also estimated via the Monte Carlo method. The resulting stellar rotation curve is presented in Figure~\ref{modelfit} as purple lines.

 \subsection{The models of dark matter halo} 

In this paper, we employed three well-known dark matter halo models to represent the cuspy dark matter halo and cored dark matter halo profiles: Navarro-Frenk-White (NFW) model \citep{Navarro1996}, Pseudo-isothermal (ISO) halo model and Burkert model \citep{Burkert_1995}. Besides, we introduced generalized NFW model \citep{Zhao_1996} to quantitatively determine the inner power-law slope of dark matter density profile.

  \subsubsection{Navarro-Frenk-White (NFW) model}

The NFW model \citep{Navarro1996} is derived from N-body simulation based on $\mathrm{\Lambda CDM}$. It has a density profile of
  
\begin{equation}
 \rho_{\mathrm{NFW}}(R)=\frac{\rho_{\rm{c}} \delta_{\rm{char}}}{(R/R_{\rm{s}})(1+R/R_{\rm{s}})^2},
\end{equation}

\noindent where $\rho_{\rm{c}}=3H^2/8\pi G$ is the present critical density, $\delta_{\mathrm{char}}=\frac{200}{3} (c^3/[\mathrm{ln}(1+c)-c/(1+c)])$ is the characteristic dimensionless density, and $R_{\rm{s}}=R_{200}/c$ is the scale radius. $R_{200}$ is the the radius within which the mean density of the halo is 200 times the present critical overdensity, and $c$ is the concentration of halo. The mass of the halo is determined by $R_{200}$ though $M_{200}=200\rho_{\mathrm{crit}}(4\pi /3)R_{200}^3$. A prominent feature of this model is that the dark matter density in the inner region steepens toward the center following $\rho \propto R^{-1}$. 

The corresponding rotation velocity of NFW model is

\begin{equation}
V_{\mathrm{NFW}}(R)=V_{200}\sqrt{\frac{\mathrm{ln}(1+cx)-cx/(1+cx)}{x[\mathrm{ln}(1+c)-c/(1+c)]}}
\end{equation}

\noindent where $V_{200}=\sqrt{GM_{200}/R_{200}}$ is the circular velocity at $R_{200}$ and $x=R/R_{200}$.

  \subsubsection{Pseudo-isothermal halo model}

The spherical pseudo-isothermal model is motivated by observations. It has a density profile of:  

\begin{equation}
\rho_{\mathrm{ISO}}(R)=\frac{\rho_0}{1+(R/R_{\rm{c}})^2},
\end{equation}

\noindent where $\rho_0$ is the central core density and $R_{\rm{c}}$ is the core radius of the halo. Toward the galactic center, the density flattens and becomes a constant core which can be described as $\rho\propto R^0$. The rotation velocity is 

\begin{equation}
\begin{split}
V_{\mathrm{ISO}}(R) &=\sqrt{4\pi G\rho_0 R_{\rm{c}}^2[1-\frac{R_{\rm{c}}}{R}\mathrm{arctan}(\frac{R}{R_{\rm{c}}})]} \\
 &= V_{\infty} \sqrt{1-\frac{R_{\rm{c}}}{R}\mathrm{arctan}(\frac{R}{R_{\rm{c}}})},
 \end{split}
\end{equation}

\noindent in which $V_{\infty}=\sqrt{4\pi G\rho_0 R_{\rm{c}}^2}$ is the asymptotic velocity at large radii.

\subsubsection{Burkert model}

The Burkert model is an empirical law used to explain the observed rotation curve of dwarf galaxies. It is characterized by central density $\rho_0$ and core radius $R_0$ as follows \citep{Burkert_1995}:

\begin{equation}
    \rho_{\mathrm{bk}}(R)=\frac{\rho_0 R_0^3}{(R+R_0)(R^2+R_0^2)}.
\end{equation}

\noindent The density decreases as $\rho \propto R^{-3}$ in the outer region. By integrating the density profile, the rotation velocity can be written as \citep{Shi-2018}:

\begin{equation}
    V_{\mathrm{bk}}(R)=V_{\rm{c}}\sqrt{\frac{2\mathrm{ln}(1+x)-2\mathrm{arctan}(x)+\mathrm{ln}(1+x^2)}{x[3\mathrm{ln}(2)-2\mathrm{arctan}(1)]}},
\end{equation}

\noindent where $x=R/R_0$ and $V_{\rm{c}}$ is the circular velocity at core radius $R_0$.

\subsubsection{gNFW model}
gNFW model is a generalized form of NFW model proposed by \cite{Zhao_1996}. It describes a density profile with an arbitrary power-law slope $-\gamma$ in the inner region and declines as $\rho \propto R^{-3}$ at radius much larger than scale radius \citep{Wyithe_2001}:

\begin{equation}
    \rho_{\mathrm{gNFW}}(R)=\frac{\rho_{\rm{c}} \delta_{\mathrm{char}}}{(\frac{R}{R_{\rm{s}}})^{\gamma}(1+\frac{R}{R_{\rm{s}}})^{3-\gamma}}.
\end{equation}

\noindent The definition of $\rho_{\rm{c}}$ and $\delta_{\rm{char}}$ are identical to those in the NFW model. The generalized form reproduces the NFW model when $\gamma=1$.

 \subsection{The Bayesian inference of dark matter halo models}
Compared with the density profile, the rotation curve carries the same information about the mass distribution while being less noisy, as it reflects the cumulative radial structure of the halo \citep{Navarro1996}. Accordingly, we employed a Bayesian statistical framework to fit the above dark matter halo models to the observed rotation curve to infer the posterior distribution of the model parameters. The Bayesian inference was performed using the \texttt{Python} package \texttt{PyMC} \citep{salvatier2015}, which implements Markov chain Monte Carlo (MCMC) sampling algorithms to explore the posterior parameter space. 

We first fitted the observed rotation curve with the dark matter halo models using the \texttt{curve\_fit} function from \texttt{Scipy} \citep{scipy2020} to obtain initial parameter estimates, which were subsequently used as the mean priors for Bayesian inference. However, the fit of NFW model yielded unphysical solutions, with the best-fit concentration $c$ taking negative value. As a result, we manually specified the prior distributions for the parameters of NFW model. During the Bayesian analysis, we adopted a truncated normal distribution with a lower bound of zero for all parameters, and set the prior standard deviations to twice the corresponding means to allow sufficient flexibility. Sampling was performed using the default No-U-Turn Sampler (NUTS) implemented in \texttt{PyMC}, with $\mathrm{draws} = 5000$, $\mathrm{tune} = 2000$, $\mathrm{chains} = 4$, and $\mathrm{target\_accept} = 0.99$. The prior for each parameter is shown in Table~\ref{model priors}.

For gNFW model which lacks an analytical solution, we used numerical integration together with \texttt{curve\_fit} to fit it to our rotation curve.

\begin{table}
\centering
\caption{Prior distributions for dark matter models in \texttt{PyMC} fits}
\label{model priors}
\renewcommand{\arraystretch}{1.5}
\begin{tabular}{c c}
 \hline
 \hline
 Parameters & Truncated Nomral ($\mu,\,\sigma$) \\
 \hline
 NFW model \\
 \hline
 $V_{200}$ & (42, 84) $\mathrm{km\,s^{-1}}$\\
 $R_{\rm{s}}$ & (150, 150) kpc \\
 \hline
 ISO model \\
 \hline
 $V_{\infty}$ & (113.87, 227.74) $\mathrm{km\,s^{-1}}$ \\
 $R_{\rm{c}}$ & (7.37, 14.74) kpc \\
 \hline
 Burkert model \\
 \hline
 $V_{\rm{c}}$ & (67.92, 135.84) $\mathrm{km\,s^{-1}}$ \\
 $R_0$ & (11.44, 22.88) kpc \\
 \hline
 \hline
\end{tabular}
\end{table}

\section{RESULTS}

 \subsection{The best-fit model of rotation curve}
\begin{figure*}
    \centering
    \includegraphics[width=0.7\linewidth]{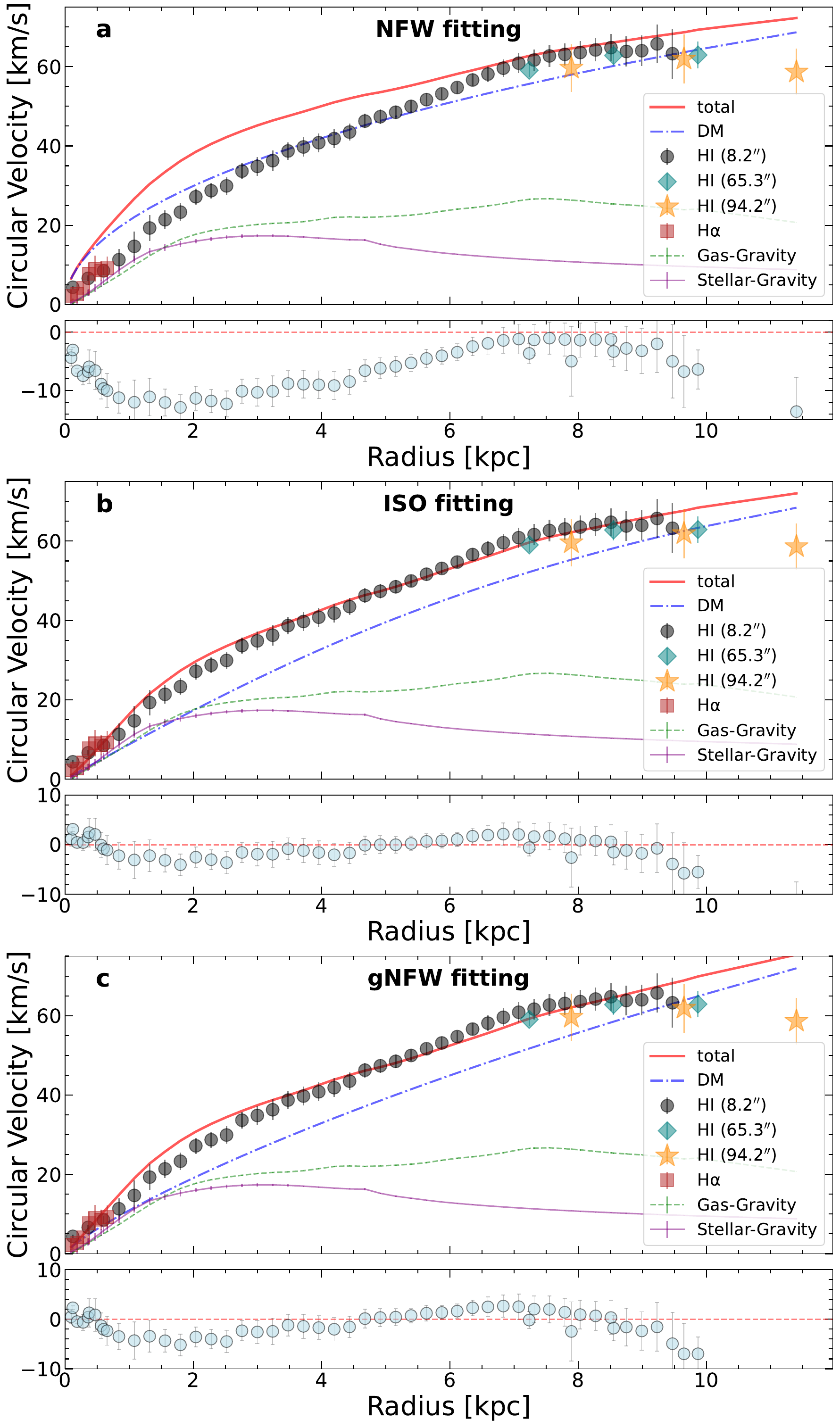}
    \caption{\texttt{PyMC} best-fit results for the NFW, ISO, and gNFW models. In each panel, the red solid and blue dashed lines show the total and dark-matter-only rotation curves, with residuals plotted in the lower subpanels. The data points are: red squares for H$\alpha$ in the inner region, black circles for \texttt{r00\_t00}, green diamonds for \texttt{r05\_t60}, and yellow stars for \texttt{r10\_t90}. The green and purple lines represent the gas and stellar contributions.}
    \label{modelfit}
\end{figure*}

All models achieved convergence during the fitting procedure (Figure~\ref{modelfit}). The NFW model overpredicts the inner velocities and gives a poor fit, with reduced chi-square $\chi^2/{\rm d.o.f.}=19.86$. By contrast, the ISO and Burkert models reproduce the observed rotation curve well, with both $\chi^2/{\rm d.o.f.}=2.04$. The ISO model yields a core radius of $7.62\pm0.80$ kpc, and the Burkert model gives $11.97\pm1.44$ kpc. Both values imply large core radii, and the difference arises because the Burkert profile drops off more rapidly in the outer region, requiring a larger central dark matter core. The detailed Bayesian inference results are presented in Table~\ref{posterior_distribution}. We derived the inner density slope of dark matter from the fit of gNFW model, which gives a best-fit slope of $\gamma=0.36\pm0.12$. Compared with the NFW model slope of $\gamma=1$, our result is quantitatively closer to a constant-density core.
 
Such a large dark matter core radius is uncommon among dwarf galaxies. The stellar surface brightness profile has an effective radius of $R_{\rm{e}}=2.02\pm0.10$ kpc, so the ISO core radius alone corresponds to about $3.8R_{\rm{e}}$. The dark matter core therefore extends far beyond the stellar body of the galaxy. 

\begin{figure*}
    \centering
    \includegraphics[width=\linewidth]{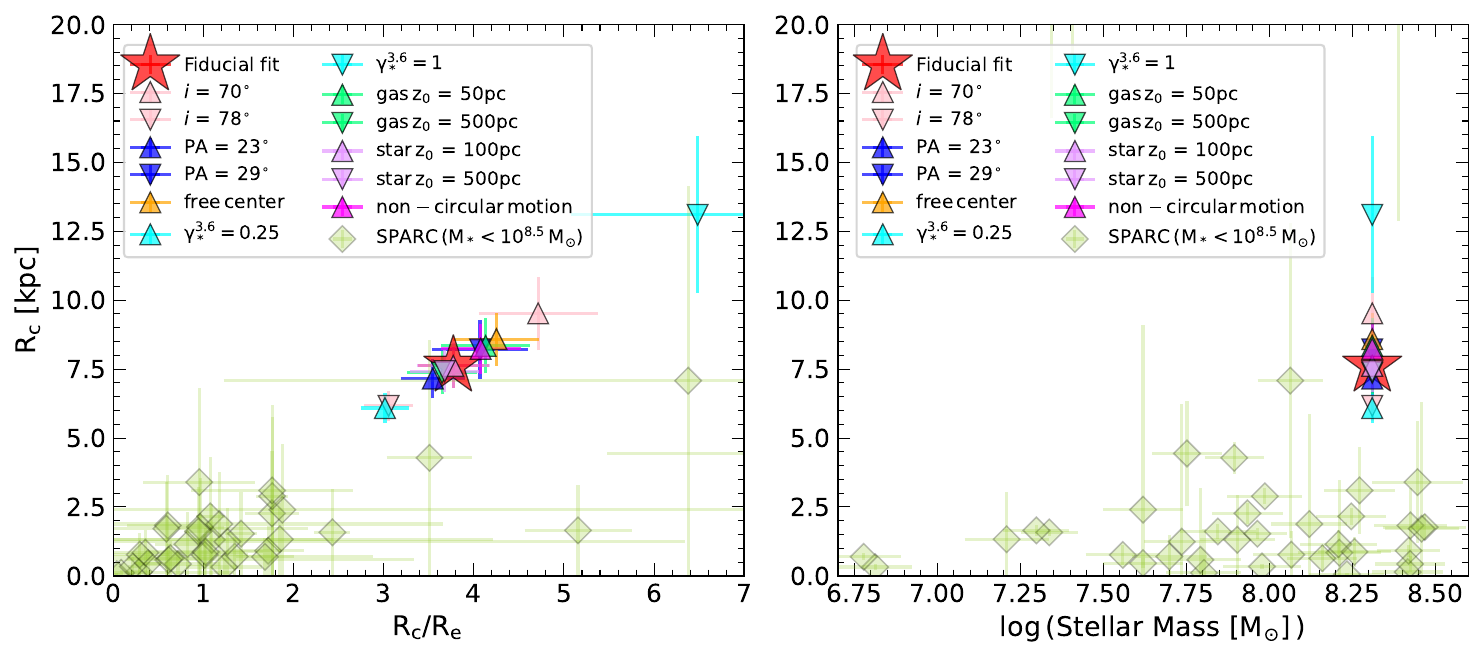}
    \caption{ISO model core radius $R_{\mathrm{c}}$ as a function of $R_{\mathrm{c}}/R_{\mathrm{e}}$ (left) and stellar mass $M_*$ (right). The red star denotes the fiducial result of this work, and the colored triangles indicate various systematic uncertainty estimates. The light green diamonds correspond to SPARC galaxies with $M_*<10^{8.5}\mathrm{M_{\odot}}$.}
    \label{core_mass}
\end{figure*}

To place UGCA 320 in context, we compare its ISO core radius with those of galaxies ($M_{*}<10^{8.5}\,\rm{M_{\odot}}$) in the SPARC database (Spitzer Photometry \& Accurate Rotation Curve; \citealt{Lelli_2016}) using the ISO model fits from \cite{Li_2020} for the SPARC galaxies. Notably, UGCA 320 is also included in SPARC as DDO 161. The HI observation was carried out using the Very Large Array (VLA) with a total integration time of four hours, resulting in a beam size of $13^{\prime\prime} \times 13^{\prime\prime}$ \citep{cote2000}. The derived rotation curve from \cite{cote2000}, however, only covers roughly half of our current radial extent, and the ISO core radius reported by \cite{Li_2020} is $3.79\pm1.45$ kpc after scaling to the newly determined TRGB distance. As shown in Figure~\ref{core_mass}, which plots $R_{\mathrm{c}}$ against both $R_{\mathrm{c}}/R_{\mathrm{e}}$ (left) and stellar mass (right), UGCA 320 stands out with a significantly larger core radius than other dwarfs---a result that remains robust after accounting for various systematic uncertainties.

 \subsection{The systematic uncertainties of the rotation curve}
We examined the systematic uncertainties that may affect the cored dark matter rotation curve and found that our results are robust against these uncertainties.

\begin{figure*}
    \centering
    \includegraphics[width=\linewidth]{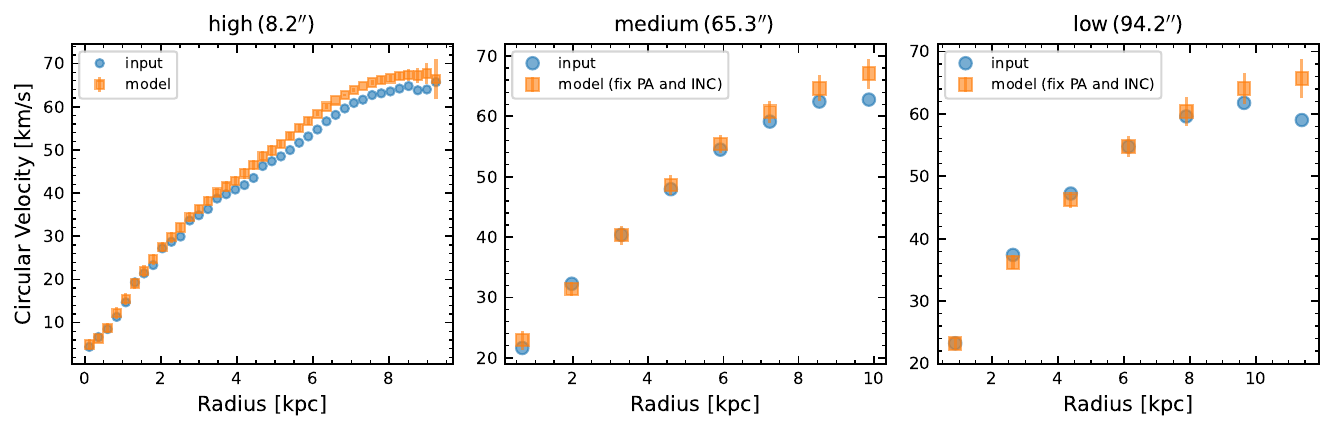}
    \caption{Comparison of the mock data cubes (orange squares) and the recovered rotation curves (blue circles) at three spatial resolutions. The mock data cubes were generated by $\rm{^{3D}BAROLO}$ \texttt{GALMOD} task with the inclination angle fixed to $74^{\circ}$. At this high inclination angle, $\rm{^{3D}BAROLO}$ successfully recovers the rotation curves at all resolutions, though small discrepancies appear in the outermost regions.}
    \label{mock}
\end{figure*}

\textbf{(1) Inclination angle:} For edge-on galaxies, determining the inclination angle is challenging because the galactic disk is not infinitely thin. Moreover, the inclination angle derived from $\rm{^{3D}BAROLO}$ depends sensitively on the initial guess at both low and high inclination angles. As noted by \cite{diteodoro2015}, galaxies with inclination angles in the range $45^{\circ}\lesssim i \lesssim 75^{\circ}$ can yield acceptable fits with $\rm{^{3D}BAROLO}$, and our galaxy lies near the upper bound. To test the ability of $\rm{^{3D}BAROLO}$ to recover the rotation curve at such high inclination angles, we used the $\rm{^{3D}BAROLO}$ \texttt{GALMOD} task to generate mock HI datacubes at three different spatial resolutions, fixing $i=74^{\circ}$ (Figure~\ref{mock}). The mock rotation curves generally agree well with the input data, although at the outermost radii the mock velocities are slightly higher. For the highest resolution ($8.2^{\prime\prime}$), the mock velocity exceeds the input one by up to $\sim3.6\,{\mathrm{km\,s^{-1}}}$, while for the lowest resolution ($94.2^{\prime\prime}$), the mean difference is about $\sim1.7\,\mathrm{km\,s^{-1}}$. Using these mock rotation curves together with the same baryonic distribution, we performed the model fits, and the fitting results are fully consistent with the fiducial fits. Therefore, we conclude that at inclination angles as high as $\sim 74^{\circ}$, $\rm{^{3D}BAROLO}$ can derive reliable rotation curves without biasing the ISO core radius (if present).

The uncertainty in inclination angle has negligible impact on the derived rotation curve for $i>75^{\circ}$, since it enters through $\mathrm{sin}i$. By contrast, the baryonic mass surface density is sensitive to inclination angle, scaling as $\mathrm{1/cos}i$. To account for this, we adopt an uncertainty of $74\pm4^{\circ}$. The corresponding results (Figure~\ref{sys_geo} a-d) indicate that a cored dark matter distribution remains better fitted to the data. At $78^{\circ}$, the ISO core radius falls to $6.17\pm0.55$ kpc, and Burkert core radius to $9.63\pm0.93$ kpc.

\textbf{(2) Position angle:} The radial profile of position angle from $\rm{^{3D}BAROLO}$ suggests an uncertainty of $\pm3^{\circ}$. As shown in Figure~\ref{sys_geo} e-h, a shift to $293^{\circ}$ reduces the ISO core radius to $7.16\pm0.72$ kpc and the Burkert core radius to $11.03\pm1.22$ kpc, and the inner density slope rises to $0.54\pm0.12$. At $299^{\circ}$, however, the derived parameters show little variation.

\textbf{(3) Kinematic center:} In our fiducial fits, the kinematic center was fixed to the 2MASS NIR photometric center. To evaluate the systematic uncertainty of this assumption, we allowed the kinematic center to vary freely. The best-fit free center is offset from the photometric center by $4.36^{\prime\prime}$ ($\sim130\,\mathrm{pc}$), well within the beam size of the highest-resolution HI cube ($8.2^{\prime\prime}$). From Figure~\ref{sys_geo} i-l, the resulting rotation curve and model fits are compatible with the fiducial ones, yielding ISO and Burkert core radii of $8.59\pm0.96$ kpc and $14.06\pm1.94$ kpc, respectively.

\textbf{(4) Mass-to-light ratio in 3.6 $\mathrm{\mu m}$:} The stellar mass is inferred from the mass-to-light ratio, introducing uncertainties that propagate into both the stellar and dark matter mass distribution. We test the robustness by varying $\gamma_{3.6\mathrm{\mu m}}$ between 0.25 and 1 and refitting the dark matter models (Figure~\ref{sys_ML_non} a-d). At $\gamma_{3.6\mathrm{\mu m}}=0.25$, the ISO core radius decreases to $6.09\pm0.55$ kpc, and the Burkert core radius drops to $9.45\pm0.88$ kpc. The inner density slope of gNFW model becomes $0.39\pm0.09$.

\begin{figure}
    \centering
    \includegraphics[width=\linewidth]{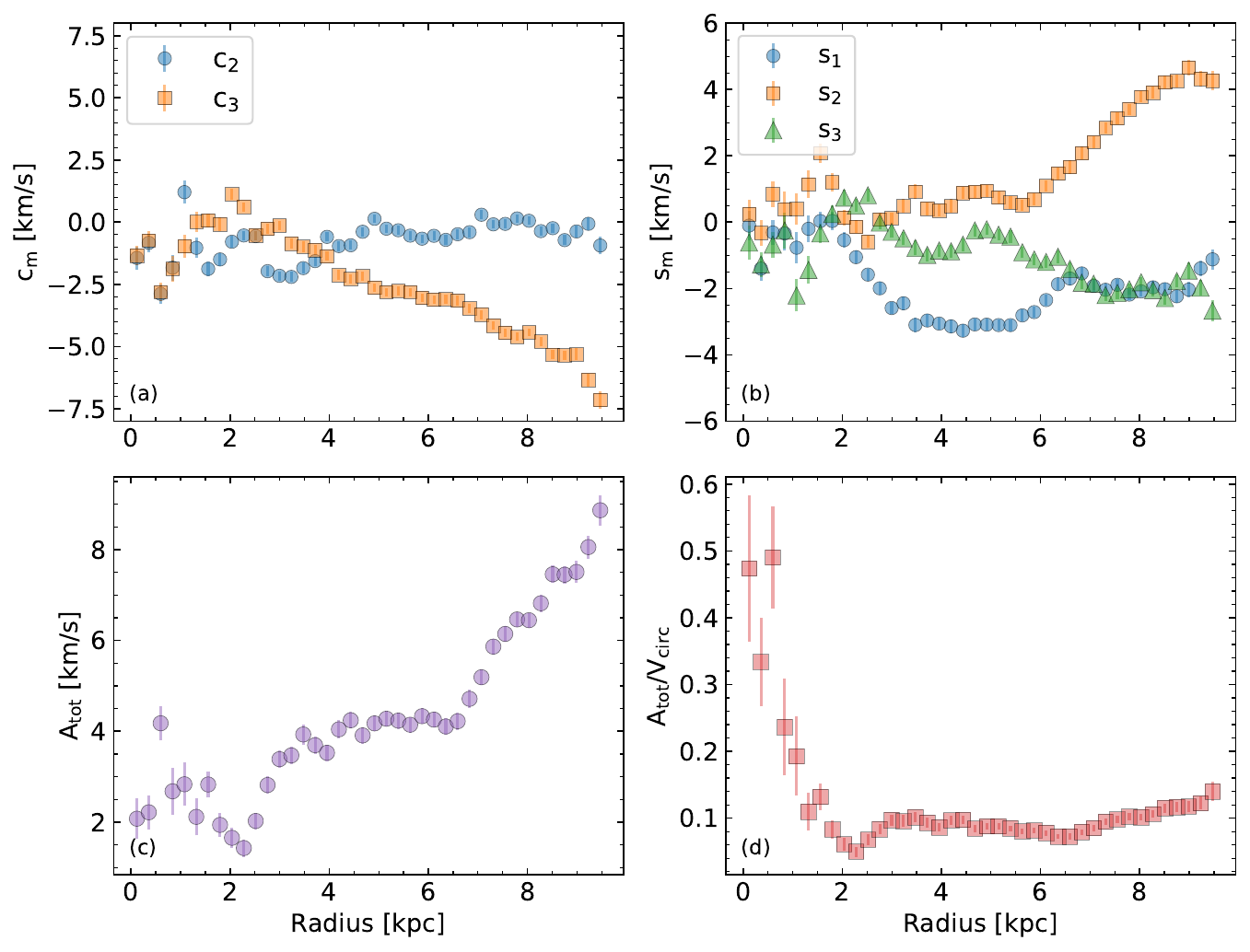}
    \caption{Harmonic decomposition of non-circular motions. Panels (a) and (b) show the radial distrubutions of the harmonic terms. Panel (c) presents the total amplitude of non-circular motions, and panel (d) displays the ratio of non-circular to circular motions.}
    \label{noncircular}
\end{figure}

\textbf{(5) Non-circular motion:} To quantify the non-circular motions, we performed a harmonic decomposition of the residual velocity field after subtracting the tilted-ring model from the observed velocity field \citep{schoenmakers1997, Trachternach2008}. The line-of-sight velocity can be written as

\begin{equation}
    V_{\mathrm{los}}(R)=V_{\mathrm{sys}}(R)+\sum_{m=1}^N[c_{m}(R)\mathrm{cos}\,m\psi + s_m(R)\mathrm{sin}\,m\psi],
\end{equation}

\noindent where $\psi$ is the azimuthal angle in the disk plane, and $c_m$ and $s_m$ are the amplitudes of $m$-th order harmonic terms. $c_1$ corresponds to rotation, while $s_1$ traces radial motion. When we restrict to $m=1$ and ignore radial motion ($s_1=0$), the equation reduces to the purely circular motion case: $V_{\mathrm{los}}(R)=V_{\mathrm{sys}}(R)+c_1\mathrm{cos}\,m\psi$. We expanded the residual field up to $m=3$, which has been shown to capture most non-circular motions \citep{Trachternach2008}. The decomposition was performed through \texttt{GIPSY} task \texttt{RESWRI} with the \texttt{CENTER}, \texttt{PA}, and \texttt{INC} fixed, as leaving them free tends to underestimate the non-circular motions \citep{Oh_2015}. Figure~\ref{noncircular} presents the resulting harmonic amplitudes. The total non-circular amplitude is

\begin{equation}
    A_{\mathrm{tot}}(R)=\sqrt{s_1^2+c_2^2+s_2^2+c_3^2+s_3^2},
\end{equation}

\noindent with a median value of $\sim4.2\,\mathrm{km\,s^{-1}}$.

Within the central 1 kpc, the non-circular amplitudes are comparable to the circular velocity. To assess their influence, we excluded the data points with $A_{\mathrm{tot}}/V_{\mathrm{circ}}>20\%$, as well as H$\alpha$ data points for which non-circular motions could not be reliably measured. As plotted in Figure~\ref{sys_ML_non} e-h, the fitting results change only marginaly. After removing the inner four HI points and H$\alpha$ data, the ISO core radius becomes $8.23\pm0.90$ kpc and Burkert core radius $13.31\pm1.82$ kpc. However, the inner density slope in this case is no longer well constrained.

\textbf{(6) The scale height of gas and stellar disk: }In our $\rm{^{3D}BAROLO}$ and \texttt{GIPSY} \texttt{ROTMOD} setups, we fixed the gas and stellar disk scale heights to 100 pc and 200 pc, respectively. We varied the gas scale height to 20 pc and 500 pc, and the stellar scale height to 100 pc and 500 pc (Figure~\ref{sys_height}). The final results remained essentially unchanged.

Among all systematic uncertainties, the smallest core radii are obtained when $\gamma_{3.6\mathrm{\mu m}}=0.25$ or $i=78^{\circ}$, yielding ISO core radii of $\sim6$ kpc and Burkert core radii of $\sim9.5$ kpc. Nevertheless, all these values remain within the $3\sigma$ uncertainty of the fiducial fits, confirming the robustness of our results. Compared with other dwarf galaxies, these core radii are still remarkably large, placing UGCA 320 in an extraordinary region of $R_{\mathrm{c}}-R_{\mathrm{c}}/R_{\mathrm{e}}$ diagram (Figure~\ref{core_mass}).

\section{Discussion}
\subsection{Dark matter candidates}
\subsubsection{Warm dark matter}
Warm dark matter (WDM) is a dark matter candidate with higher velocity and lower mass than cold dark matter (CDM). In the WDM scenario, WDM decouples from radiation in the early universe while retaining a non-zero thermal velocity. On the one hand, this leads the WDM particles to move freely from the high-density regions to low-density regions, damping the mass fluctuations on small scales — an effect known as free-streaming. All the primordial perturbations smaller than the free-streaming scale $\lambda_{fs}$ are erased \citep{Bode_2001,Macci__2012}, which helps to explain the missing satellites problem. On the other hand, according to Tremaine-Gunn limit \citep{PhysRevLett.42.407}, the maximum phase space density is finite. A coarse-grained or pseudo phase space density,
\begin{equation}
    Q\equiv \frac{\rho}{\sigma^3},
\end{equation}

\noindent is often used as a proxy for the fine-grained phase space density \citep{Hogan_2000}. Since SIDM has a non-zero thermal velocity dispersion, $Q_{\rm{max}}$ of WDM is much smaller than that of CDM, which limits the central dark matter density to increase sharply. Consequently, in the WDM scenario, dark matter exhibits a cored density distribution in the center.

 The thermal velocity can be computed as \citep{Bode_2001}:

\begin{equation}
    \frac{v_0(z)}{1+z}=.012(\frac{\Omega_{\nu}}{0.3})^{\frac{1}{3}}(\frac{h}{0.65})^{\frac{2}{3}}(\frac{1.5}{g_{\nu}})^{\frac{1}{3}}(\frac{\mathrm{keV}}{m_{\nu}})^{\frac{4}{3}}\,\,\mathrm{km\,s^{-1}},
\end{equation}

\noindent in which $m_{\nu}$ is WDM particle mass and $g_{\nu}$ is the degree of freedom, for which we use the value of light neutrino for reference. For a Fermi-Dirac distribution, the rms velocity $\sigma=3.571\times v_0$. Combining these, we can derive

\begin{equation}
    Q_{\mathrm{max}}=1.64\times 10^{-3}(\frac{\rho_{\mathrm{L}}}{\rho_{\mathrm{cr}}})(\frac{m_{\nu}}{\mathrm{keV}})^4 \frac{\mathrm{M_{\odot}\,pc^{-3}}}{\mathrm{(km\,s^{-1})^3}},
\end{equation}

\noindent where $\rho_{\mathrm{L}}$ and $\rho_{\mathrm{cr}}$ are the local density and critical density of the Universe.

The dark matter core size can be approximated by \cite{Hogan_2000}:

\begin{equation}
    r_{\mathrm{core}}^2=\frac{\sqrt{3}}{4\pi G Q_{\mathrm{max}}}\frac{1}{<\sigma^2>^{1/2}}.
\end{equation}

With the equations above, we can estimate the mass of WDM approximately. We use the asymptotic velocity $V_{\infty}/\sqrt{2}$ and the central core density $\rho_0$ from ISO model to represent $\sigma$ and $\rho_L$, respectively. The best-fit parameters derived from ISO model leads to a WDM particle mass $\sim0.003$ keV.

However, this mass is inconsistent with current research. \cite{Enzi_2021} combined strong gravitational lensing, Lyman-$\alpha$ forest, and Milky Way satellite galaxies counts to place a lower limit on the WDM particle mass of $m_{\rm{WDM}}>6.048$ keV at 95\% confidence level. Recent independent analyses based solely on Lyman-$\alpha$ forest, Milky Way satellites counts, UV Luminosity functions, gravitational lensing excluded WDM mass below 5.7 keV, 6.6 keV, 3.2 keV, and 6.1 keV, respectively \citep{Ir_i__2024, Tan_2025, liu2024newgalaxyuvluminosity, keeley2024jwstlensedquasardark}. An alternative scenario is that the dark matter consists of both CDM and WDM \citep{Anderhalden_2012, Tan_2025}, in which case WDM particle mass can be smaller as the warm dark matter fraction $f_{\rm{WDM}}$ decreases. For $f_{\rm{WDM}}<0.45$, the lower limit on $m_{\rm{WDM}}$ is approximately 1.5 keV. 

\subsubsection{Self-interacting dark matter}
Unlike collisionless CDM which has only gravitational interaction, \cite{Spergel_2000} proposed dark matter particles are self-interacting with a large scattering cross section. Self-interacting dark matter (SIDM) can transmit kinematic energy from the outer part of halo inward through elastic scatterings, so that the central densities can be reduced\citep{Rocha_2013,Kaplinghat_2016,Tulin_2018,Adhikari_2025}. \cite{Miralda_Escude_2002} utilized cluster lensing to test the hypothesis of SIDM, and found that SIDM with a cross section independent of velocity is too weak to explain both the cusp-core problem and missing satellite problem. After that, SIDM with velocity-dependent cross section became popular, which has a larger cross section on galaxy scales while a smaller one on cluster scales, and is demonstrated as an exhaustive solution for astrophysical systems at different scales \citep{Zavala_2013,Sagunski_2021,Outmezguine_2023}.  

In the high-density inner region, each particle is assumed to scatter with other particles at least one time during the halo age $t_{age}$ on average \citep{Kaplinghat_2016,Tulin_2018}. Thus we have

\begin{equation}
    R_{\mathrm{scat}}(r)\times t_{\mathrm{age}} \approx \rho_{\mathrm{DM}}(r)(\sigma/m)v_{\mathrm{rms}}\times t_{\mathrm{age}}\approx 1,
\end{equation}

\noindent where $R_{\mathrm{scat}}$ is the scattering rate, $\sigma/m$ is the SIDM cross section per particle mass, and $v_{\mathrm{rms}}$ is the relative velocity between dark matter particles.

Assuming a typical dwarf galaxy dark matter density $\rho_{\mathrm{DM}}=0.01\,\mathrm{M_{\odot}/pc^3}$ and $v_{\mathrm{rms}}=50\,\mathrm{km\,s^{-1}}$ from THINGS dwarf galaxies \citep{Oh_2011}, and $t_{\mathrm{age}}=10$ Gyr, the equation can be written as:

\begin{equation}
    \frac{\sigma/m}{1\,\mathrm{cm^2/g}}=(\frac{10\,\mathrm{Gyr}}{t_{\mathrm{age}}})(\frac{0.01\,\mathrm{M_{\odot}/pc^3}}{\rho_{\mathrm{DM}}})(\frac{50\,\mathrm{km\,s^{-1}}}{v_{\mathrm{rms}}}).
\end{equation}

At larger radii, the dark matter density decreases and the scattering rate is negligible. Thus the density profile of SIDM is similar to that of CDM in the outer region. 

We used the dark matter density and velocity at core radius of UGCA 320 to estimate the cross section per mass is $\sim 4.27\,\mathrm{cm^2\,g^{-1}}$. \cite{Elbert_2015} figured that the largest and lowest density cores develop for cross section per mass in the range of $\sigma/m \sim 5-10\,\mathrm{cm^2\,g^{-1}}$ for local field dwarfs. \cite{Ren_2019} used an analytical model to fit the rotation curves of 135 SPARC galaxies, showing that $\sigma/m>3\,\mathrm{cm^2\,g^{-1}}$ at about 100 $\rm{km\,s^{-1}}$ is needed. The fitting to six ultra-faint dwarf galaxies yielded a wide allowed range $0.3<\sigma/m<199\,\mathrm{cm^2\,g^{-1}}$ \citep{S_nchez_Almeida_2025}. Thus our result is consistent with current researches on galaxy scales.

\subsubsection{Fuzzy dark matter}
\cite{Hu_2000} proposed an ultralight boson characterized by a single parameter--- mass of $m_{\psi} \sim 10^{-22}\,\mathrm{eV}$. According to quantum mechanics,  such a particle has a de Broglie wavelength as large as galactic scales. This de Broglie wavelength serves as a Jeans length: below it, quantum pressure counteracts gravitational collapse, leading to the formation of a stable soliton core surrounded by an NFW halo\citep{Hu_2000, Schive_2014, Mocz_2017, Burkert_2020}. This so-called fuzzy dark matter (FDM) particle is considered a viable dark matter candidate, with its prominent wave-like properties suppressing the formation of small-scale structure and producing a soliton core in the center \citep{eberhardt2025ultralightfuzzydarkmatter, Schive_2026}.

From high resolution simulation, \cite{Schive_2014} found a stable soliton solution to Schr\"odinger-Poisson equation for central density profile. Within 3$R_c$, the soliton density profile can be well approximated by

\begin{equation}
    \rho_{\mathrm{c}} \approx\frac{1.9(m_{\psi} / 10^{-23}\mathrm{eV})^{-2}(R_{\mathrm{c}}/\mathrm{kpc})^{-4}}{[1+9.1\times 10^{-2}(R/R_{\mathrm{c}})^2]^8} \mathrm{M_{\odot}\,pc^{-3}},
\end{equation}

\noindent where $m_{\psi}$ is the mass of FDM. We integrated the density profile to calculate the mass of dark matter halo and corresponding circular velocity, and then fitted it to our observation. The best-fit result yields $R_{\mathrm{c}}=7.42\pm0.75$ kpc and $m_{\psi}=(0.38\pm0.07) \times 10^{-23}\,\mathrm{eV}$. According to Fig.10 in \cite{eberhardt2025ultralightfuzzydarkmatter}, which presents the current observational constraints on FDM, this low mass is overwhelmingly excluded by various methods.

\subsection{Baryonic physics}
Apart from modifying the nature of dark matter, baryonic processes can also mitigate the discrepancies between simulations and observations in a $\Lambda$CDM framework. In simulations with a high gas density threshold for star formation, gas can dominate the gravitational potential \citep{Ben_tez_Llambay_2019}. Repeated bursts of star formation drive a large fraction of gas outflows via supernova explosions, leading to fluctuations in the central gravitational potential. If the busrt timescale is shorter than the local dynamical timescale, the non-adiabatic gravitational potential fluctuations can heat the orbits of dark matter particles, thereby flattening the inner dark matter density profile \citep{Navarro_1996, Governato_2010, Pontzen_2012, Teyssier_2013, Tulin_2018, Lazar_2020, Collins_2022}. When the gas cools down, it may be re-accreted onto the galaxy, sustaining the supernova-driven outflows. Such multiple episodes are more efficient in core formation \citep{sales2022baryonicsolutionschallengescosmological}. 

\cite{Di_Cintio_2013} simulated a suite of 31 galaxies using smooth particle hydrodynamics (SPH) as part of the Making Galaxies in a Cosmological Context (MaGICC; \citealt{Stinson_2012}) program, which includes early stellar feedback (i.e. energy from massive stars), to investigate the effects of baryonic feedback on the central dark matter density profile. The galaxy sample spans a stellar mass range of $5\times10^{5}-8.3\times10^{10}\,\mathrm{M_{\odot}}$. They found that the inner slope of dark matter density profile sensitively depends on the stellar-to-halo mass ratio. The most cored DM halo occurs at $M_*/M_{\mathrm{halo}} \sim 5\times10^{-3}$, with a corresponding circular velocity of $\sim 50\,\mathrm{km\,s^{-1}}$. Below this value, galaxies form too few stars, and the energy from supernova explosions is insufficient to produce a feedback-induced core. Above this value, stars deepen the gravitational potential well, thereby suppressing the feedback and retaining a cuspy profile \citep{Di_Cintio_2013, Lazar_2020, sales2022baryonicsolutionschallengescosmological}. 

These results have been further confirmed by other zoom-in cosmological simulations, such as FIRE (Feedback In Realistic Environments, \citealt{Hopkins_2014}) \citep{Chan_2015, O_orbe_2015, Lazar_2020, Straight_2025} and NIHAO (Numerical Investigation of a Hundred Astrophysical Objects, \citealt{Wang_2015}) \citep{Tollet_2016, Macci__2020}. \cite{Chan_2015} and \cite{O_orbe_2015} showed that only galaxies with sustained late-time star formation are able to create stable cores; otherwise, dark matter accretion and mergers would erase the them and regenerate cuspier profiles. \cite{Macci__2020} explored the halo response on the mass scales beyond $10^{12}\,M_{\odot}$ using the extended NIHAO suite, which includes AGN feedback from massive stars \citep{Blank_2019}, and found that AGN-driven feedback can partially offset the dark matter contraction induced by the central stellar components in massive halos. From observations, \cite{Read_2019} and \cite{hayashi2026cusptocoretransitiondarkmatter} found that this dependence also holds for Milky Way dwarf galaxies and SPARC galaxies.

For UGCA 320, the best-fit ISO and Burkert models yield halo masses of $M_{200}\sim 10^{11.65}\,\rm{M_{\odot}}$ and $M_{200}\sim 10^{10.46}\,\rm{M_{\odot}}$, corresponding to stellar-to-halo mass ratios of $M_*/M_{200} \sim 4.6\times 10^{-4}$ and $M_*/M_{200}\sim 7.1\times 10^{-3}$, respectively. Although the halo mass estimates are subject to large uncertainties, the inferred ratios may be within the range that allows efficient core formation via stellar feedback. Nevertheless, the typical sizes of feedback-generated cores remain a matter of debate \citep{Tulin_2018, sales2022baryonicsolutionschallengescosmological}. Previous studies suggest that feedback-induced core radii are generally smaller than $0.1-2$ kpc in dwarf galaxies \citep{Oman_2015, Tulin_2018, Teyssier_2013, O_orbe_2015, Tollet_2016, Lazar_2020, Jahn_2023, sales2022baryonicsolutionschallengescosmological}, making it challenging to account for the extremely large core observed in UGCA 320.

An alternative route is provided by dynamical friction, through which infalling baryonic clumps transfer their orbital energy and angular momentum to dark matter, thereby reducing the central dark matter density \citep{El_Zant_2001,El_Zant_2004}. The efficiency of this process depends on the properties of both the infalling clumps and the host halos, particularly the clump-to-halo mass ratio \citep{El_Zant_2001,Cole_2011}, and 
is found to be more effective in smaller halos at higher redshifts \citep{Del_Popolo_2016,delpopolo2022}. Notably, dynamical friction can flatten the central dark matter cusp prior to star formation and stellar feedback, and operate across a wide mass range---from dawrf galaxies to galaxy clusters \citep{El_Zant_2004,Del_Popolo_2009,Popolo_2014}. Recent simulations have demonstrated its potential importance in high-redshift massive galaxies \citep{Ogiya_2022}, however, existing studies of nearby dwarf and massive galaxies generally predict core sizes of sub-kpc to $\sim 2$ kpc \citep{Romano_D_az_2008,Goerdt_2010,Nipoti_2014,delpopolo2022}, which still falls short of explaining the large core seen in UGCA 320.

\section{Conclusions}
We have performed a comprehensive gas kinematic modelling study of the nearby dwarf galaxy UGCA 320, combining high-resolution H$\alpha$ data with deep, high-sensitivity HI observations from the MHONGOOSE survey. To mitigate the effects of non-circular motions, we subtract the additional HI component that deviates from circular motion in the HI spectra. Our main conclusions are summarized as follows:

(1) We fit a 3D tilted-ring model to the HI data cube and extract the H$\alpha$ rotation curve along the major axis. The hybrid rotation curve is fitted with four dark matter halo models. The cored ISO and Burkert models provide equally good fits to the observed curve. Consistently, the inner density slope inferred from the gNFW model is $0.36\pm0.12$, quantitatively supporting a cored dark matter distribution in UGCA 320.

(2) The best-fit ISO and Burkert models yield dark matter core radii of $7.62\pm0.80$ kpc and $11.97\pm1.44$ kpc, respectively. A comparison with dwarf galaxies ($M_{*}<10^{8.5}\,\rm{M_{\odot}}$) in the SPARC database shows that the ISO core radius of UGCA 320 is significantly larger than those typically found in dwarf galaxies.

(3) We examine the systematic uncertainties that could affect our fitting results. The most conservative assumptions---adopting an inclination angle of $\sim78^{\circ}$ or a stellar mass-to-light ratio to $\sim0.25$---yield the smallest core radii, reducing the ISO and Burkert core radii to $\sim 6$ kpc and $\sim 9.5$ kpc, respectively. Even in these extreme scenarios, both values remain within the 3$\sigma$ uncertainties of the fiducial results.

(4) The unusually large dark matter core inferred for UGCA 320 may pose a challenge to the CDM paradigm and motivates consideration of alternative dark matter models. Our results are broadly consistent with the SIDM scenario, while appearing to be in tension with WDM and FDM predictions. On the baryonic side, existing core-formation mechanisms, including stellar feedback and dynamical friction, have not yet been demonstrated to produce cores of comparable sizes in dwarf galaxies, either observationally or in simulations.

\section*{Acknowledgements}

This work is based on observations with the MeerKAT radio telescope. The MeerKAT telescope is operated by the South African Radio Astronomy Observatory (SARAO), which is a facility of the National Research Foundation, an agency of the Department of Science and Innovation. This work has made use of data from the MHONGOOSE survey (MeerKAT HI Observations of Nearby Galactic Objects: Observing Southern Emitters). We thank Dr. E. de Blok for providing the highest-resolution \texttt{r00\_t00} cube, which is beyond the public data release.

Based on observations collected at the European Southern Observatory under ESO programme: 105.20GY.002.

Based on data acquired on the ANU 2.3-metre telescope, under program: 2513051. The automation of the telescope was made possible through an initial grant provided by the Centre of Gravitational Astrophysics and the Research School of Astronomy and Astrophysics at the Australian National University and through a grant provided by the Australian Research Council through LE230100063. The Lens proposal system is maintained by the AAO Research Data \& Software team as part of the Data Central Science Platform. We acknowledge the traditional custodians of the land on which the telescope stands, the Gamilaraay people, and pay our respects to elders past and present.

The Pan-STARRS1 Surveys (PS1) and the PS1 public science archive have been made possible through contributions by the Institute for Astronomy, the University of Hawaii, the Pan-STARRS Project Office, the Max-Planck Society and its participating institutes, the Max Planck Institute for Astronomy, Heidelberg and the Max Planck Institute for Extraterrestrial Physics, Garching, The Johns Hopkins University, Durham University, the University of Edinburgh, the Queen's University Belfast, the Harvard-Smithsonian Center for Astrophysics, the Las Cumbres Observatory Global Telescope Network Incorporated, the National Central University of Taiwan, the Space Telescope Science Institute, the National Aeronautics and Space Administration under Grant No. NNX08AR22G issued through the Planetary Science Division of the NASA Science Mission Directorate, the National Science Foundation Grant No. AST–1238877, the University of Maryland, Eotvos Lorand University (ELTE), the Los Alamos National Laboratory, and the Gordon and Betty Moore Foundation.

\section*{Data Availability}

The MHONGOOSE data used in this work are available at the MHONGOOSE public data release (https://mhongoose.astron.nl/) under the target name J1303-17b. The highest-resolution \texttt{r00\_t00} cube is not included in the public release, but it is avaiable upon request. The MUSE and WiFeS data are available from the corresponding author upon reasonable request. The IRAC data are accessible via the NASA/IPAC Infrared Science Archive (https://irsa.ipac.caltech.edu/data/SPITZER/LVL/).



\bibliographystyle{mnras}
\bibliography{main} 



\appendix

\section*{Appendix}

\begin{table*}
    \centering
    \begin{tabular}{l l}
    \hline
    Parameter & Value\\
    \hline
    Enable a 3D fitting algorithm....................................[3DFIT]  & TRUE\\
    The separation between rings................................[RADSEP]  & 8.2 [$^{\prime\prime}$]\\
    The number of rings used to fit..............................[NRADII] & 40\\
    Systematic velocity......................................................[VSYS] & 739.9 [km/s]\\
    X-center of the rings...................................................[XPOS] & 13:03:16.74\\
    Y-center of the rings...................................................[YPOS] & -17:25:22.9\\
    Rotation velocity........................................................[VROT] & 36.71 [km/s]\\
    Velocity dispersion.....................................................[VDISP] & 10.95 [km/s]\\
    Inclination angle.............................................................[INC] & 74 [$^{\circ}$]\\
    Position angle...................................................................[PA] & 296 [$^{\circ}$]\\
    Scale height of the disc......................................................[Z0] & 3.44 [$^{\prime\prime}$]\\
    Distance of the galaxy.........................................[DISTANCE] & 6.03 [Mpc]\\
    Redshift of the galaxy.........................................[REDSHIFT] & 0.0025\\
    Free parameters...........................................................[FREE] & VROT VDISP PA INC VSYS\\
    Type of mask..............................................................[MASK] & SEARCH\\
    Side of the galaxy to be fitted.......................................[SIDE] & B\\
    Type of normalization................................................[NORM] & AZIM\\
    Layer type along Z-axis.............................................[LTYPE] & 1\\
    Function to be minimized.........................................[FTYPE] & 1\\
    Weighting function..................................................[WFUNC] & 0\\
    Exponent of weight for blank pixels....................[BWEIGHT] & 1\\
    Minimization to the tolerance.......................................[TOL] & 0.001\\
    Enable the second fitting stage........................[TWOSTAGE] & TRUE\\
    Type of regularisation for second fitting stage....[REGTYPE] & BEZIER\\
    Estimate the error........................................[FLAGERRORS] & TRUE\\
    Calculate the asymmetric drift correction..............[ADRIFT] & TRUE\\
    Normalize the input cube before fitting......[NORMALCUBE] & TRUE\\
    \hline
    \end{tabular}
    \caption{$\rm{^{3D}BAROLO}$ initial parameters for \texttt{r00\_t00} cube}
    \label{initpara}
\end{table*}

\begin{figure*}
    \centering
    \includegraphics[width=\linewidth]{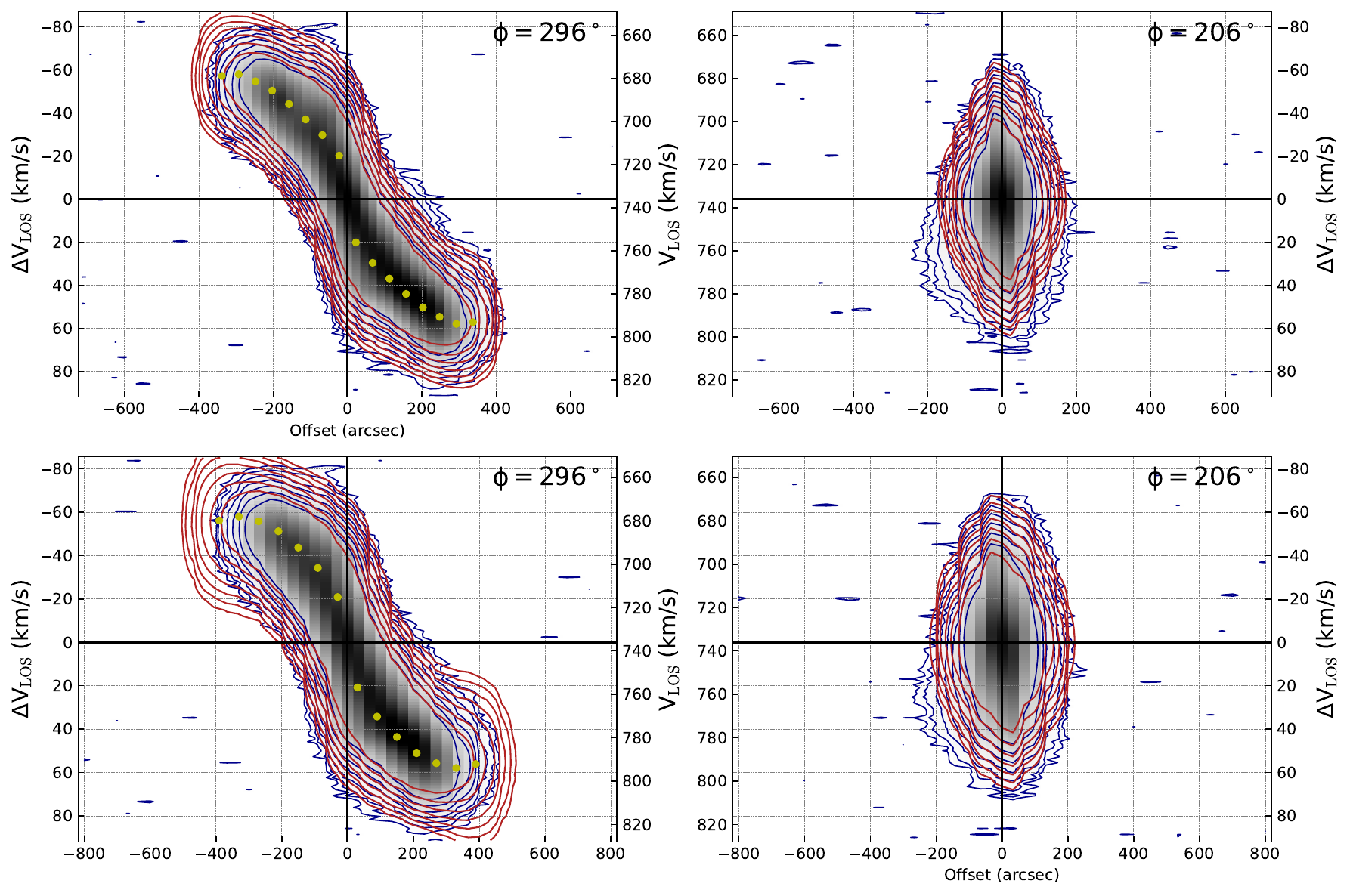}
    \caption{Position-velocity (PV) diagrams along the major axis of the \texttt{r50\_t60} (upper panel) and \texttt{r10\_t90} cubes (lower panel). The overlaid blue and red contours represent the observed and the $\rm{^{3D}BAROLO}$ modelled HI intensity, respectively. The yellow circles indicate the line-of-sight velocity, and the position angle is given in the upper-right corner.}
    \label{pv_med_low}
\end{figure*}

\begin{figure*}
    \centering
    \includegraphics[width=0.75\linewidth]{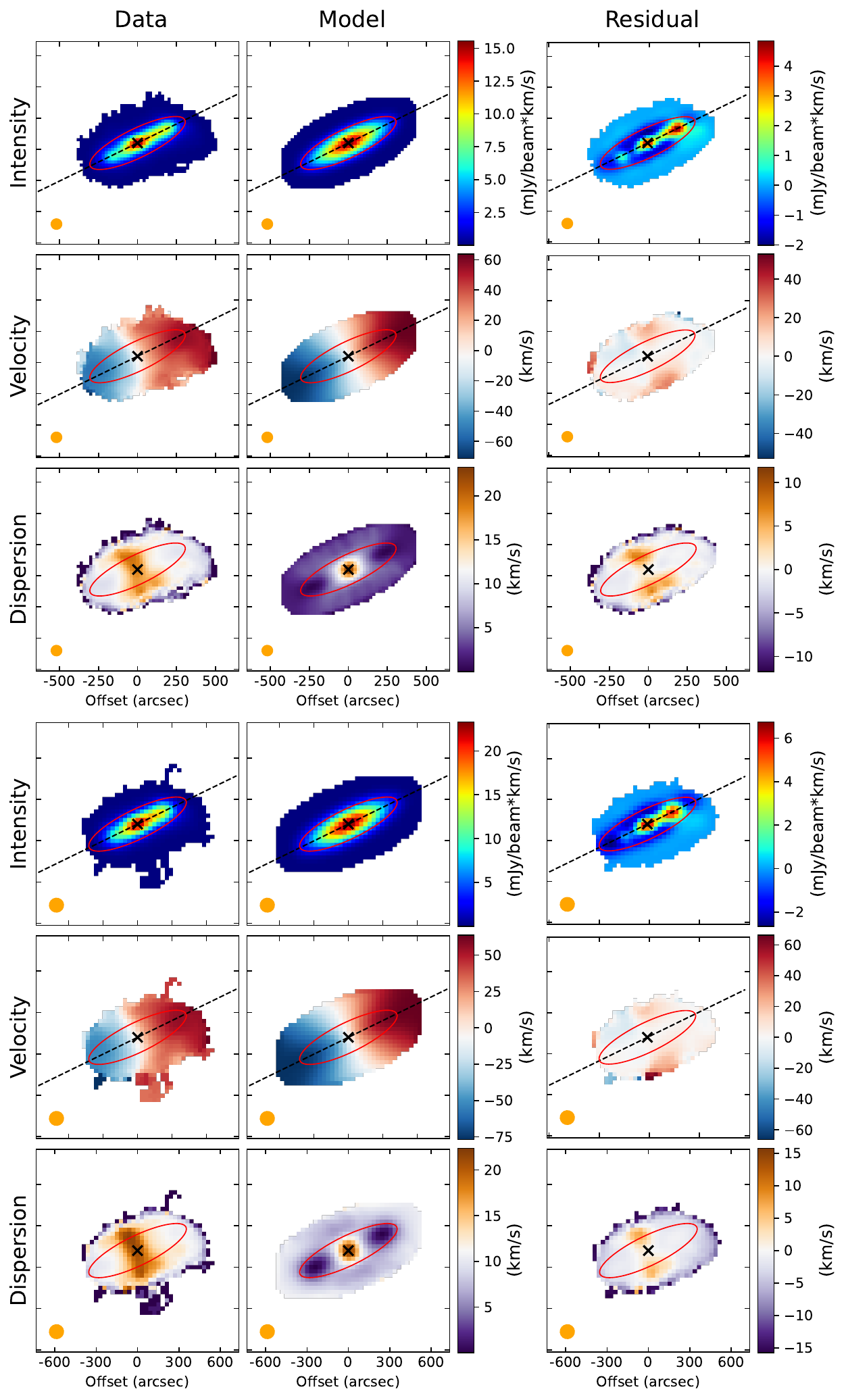}
    \caption{Moment maps of the \texttt{r05\_t60} (upper panel) and \texttt{r10\_t90} (lower panel) cubes, comparing the observations with the $\rm{^{3D}BAROLO}$ models. The corresponding beam sizes are $\mathrm{65.3^{\prime\prime}\times 64.0^{\prime\prime}}$ and $\mathrm{94.2^{\prime\prime}\times 91.2^{\prime\prime}}$, respectively.}
    \label{3DB_med_mom&low}
\end{figure*}

\begin{figure*}
    \centering
    \includegraphics[width=0.85\linewidth]{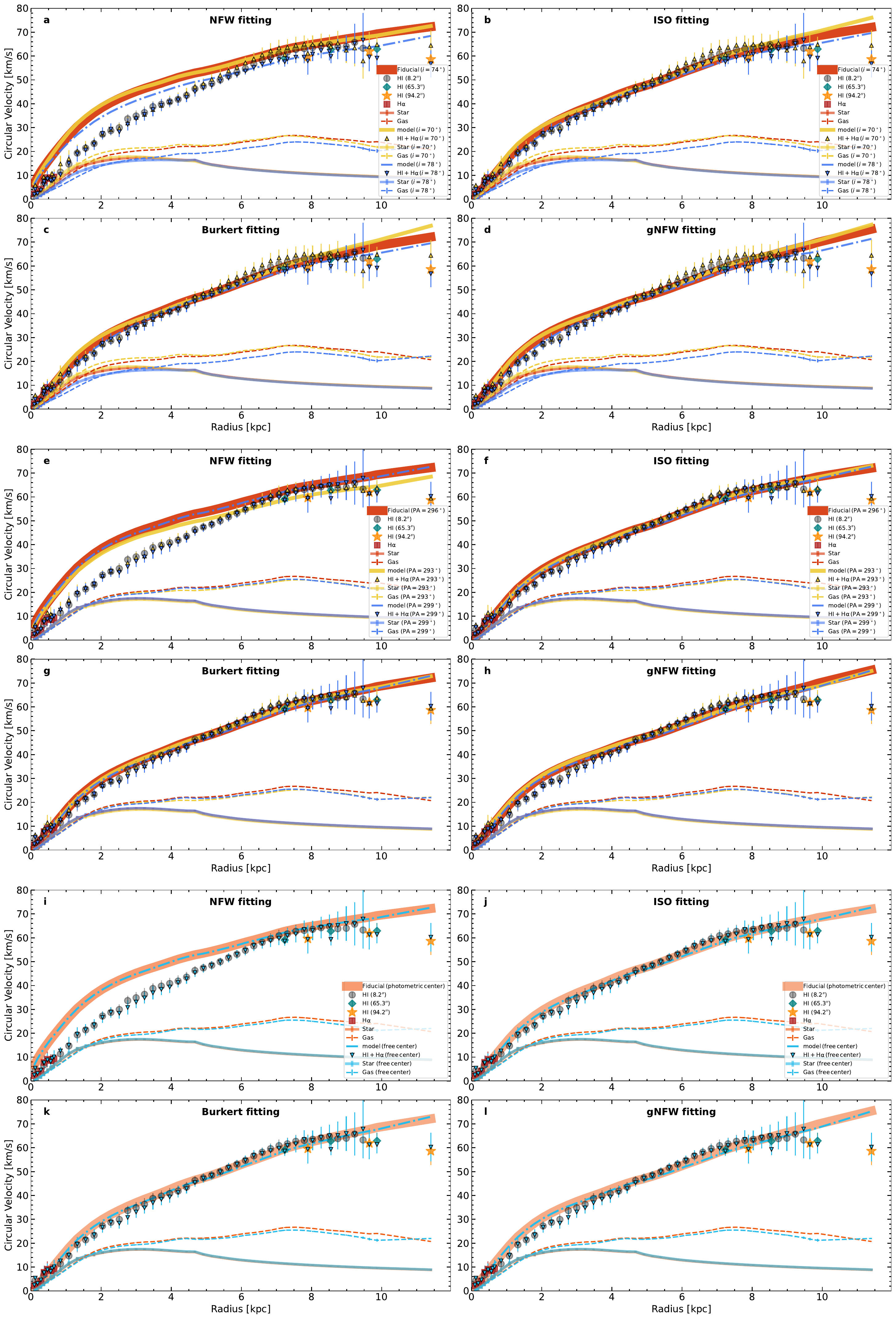}
    \caption{Systematic uncertainties in the inclination angle, position angle, and kinematic center. a-d, Results for tests with $i\pm4^{\circ}$. e-h, Results for tests with $\rm{PA}\pm3^{\circ}$. i-l, Results with the kinematic center allowed to vary freely. The thickest line corresponds to the fiducial fit and other symbols are defined in the legend.}
    \label{sys_geo}
\end{figure*}

\begin{figure*}
    \centering
    \includegraphics[width=\linewidth]{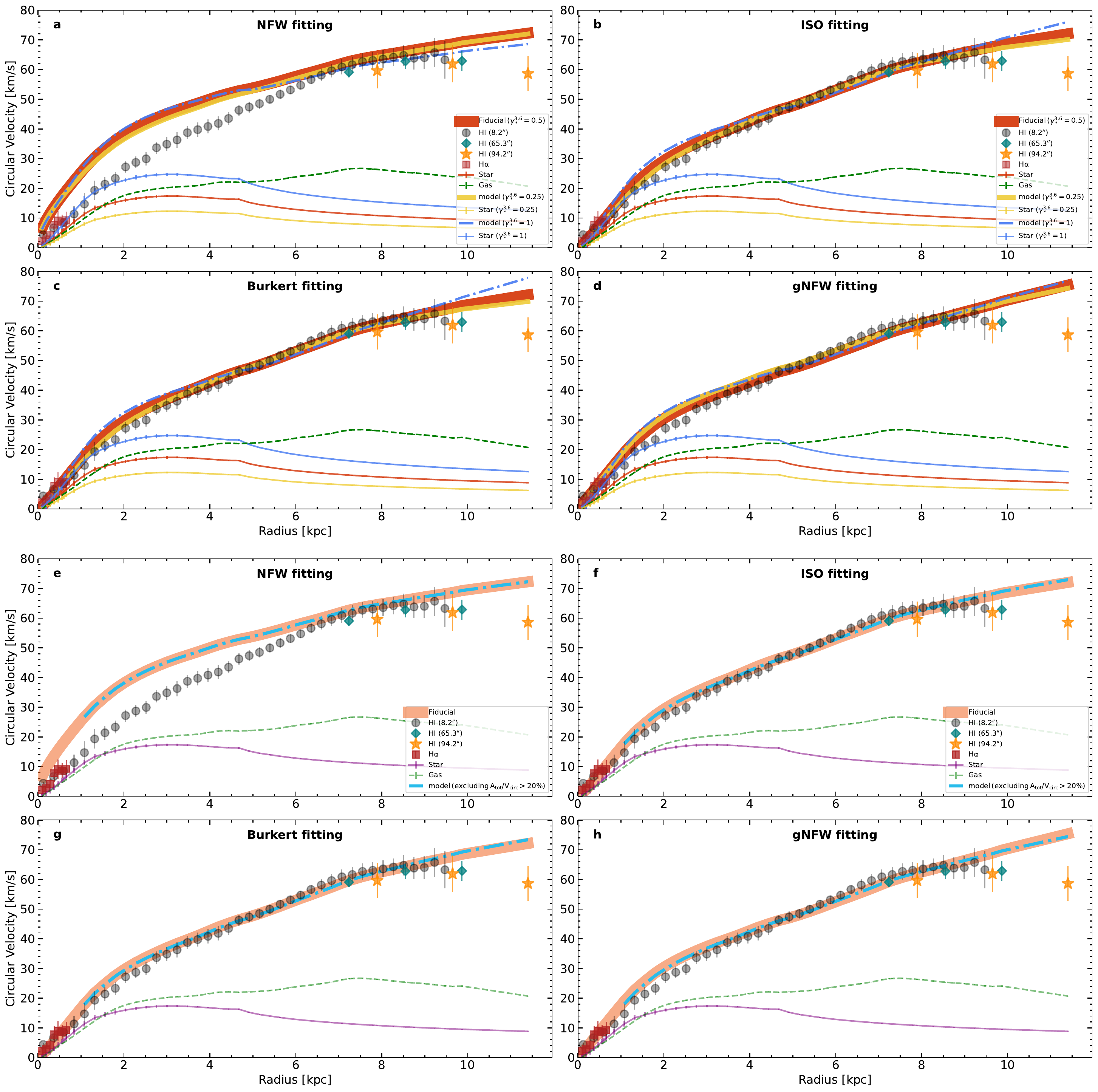}
    \caption{Systematic uncertainties in the stellar mass-to-light ratio and non-circular motions. a-d, Results for stellar mass-to-light ratio varied to 0.25 and 1. e-h, Results after excluding the H$\alpha$ and HI data points with $A_{\rm{tot}}/V_{\rm{circ}}>20\%$, where $A_{\rm{tot}}$ is the total amplitude of non-circular motions. The thickest line corresponds to the fiducial fit and other symbols are defined in the legend.}
    \label{sys_ML_non}
\end{figure*}

\begin{figure*}
    \centering
    \includegraphics[width=\linewidth]{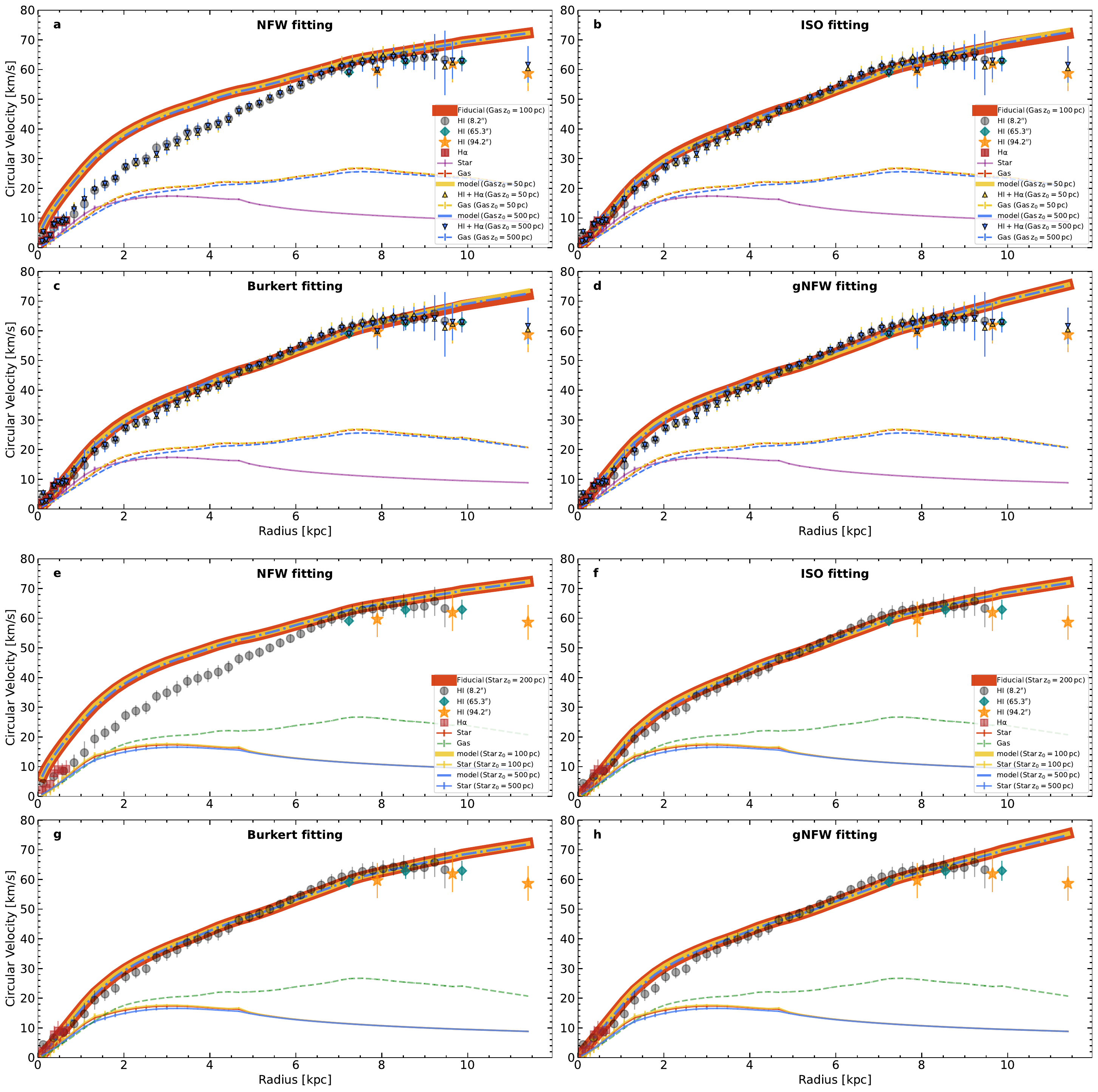}
    \caption{Systematic uncertainties in the gas and stellar scale heights. a-d, Results for gas scale height varied to 50 pc and 500 pc. e-h, Results for stellar scale height varied to 100 pc and 500 pc. The thickest line corresponds to the fiducial fit and other symbols are defined in the legend.}
    \label{sys_height}
\end{figure*}

\begin{landscape}
\begin{table}
\tiny
\centering
\caption{\texttt{PyMC} best-fit results for dark matter models}
\label{posterior_distribution}
\renewcommand{\arraystretch}{1.5}
\begin{tabular}{c c c c c c c c c c c c c c}
 \hline
 \hline
  & \multicolumn{3}{c}{NFW model} & \multicolumn{3}{c}{ISO model} & \multicolumn{3}{c}{Burkert model} & \multicolumn{4}{c}{gNFW model}\\
 \hline
  & $V_{200}$ & $R_{\rm{s}}$ & $\chi^2/d.o.f$ & $V_{\infty}$ & $R_{\rm{c}}$ & $\chi^2/d.o.f$ & $V_{\rm{c}}$ & $R_0$ & $\chi^2/d.o.f$ & $\rho_{\rm{s}}$ & $R_{\rm{s}}$ & $\gamma$ & $\chi^2/d.o.f$\\
  & ($\rm{km\,s^{-1}}$) & (kpc) & & ($\rm{km\,s^{-1}}$) & (kpc) & & ($\rm{km\,s^{-1}}$) & (kpc) & & ($\rm{M_{\odot}/pc^3}$) & (kpc) & & \\
 \hline
 Fiducial & $42.01\pm0.01$ & $149.94\pm 0.06$ & 19.86& $116.66\pm 8.89$ & $7.62\pm 0.80$ & 2.04 & $70.17\pm 6.12$ & $11.97\pm 1.44$ & 2.04 & $0.00118\pm 0.00128$ & $86.37\pm 129.27$ & $0.36\pm 0.12$ & 1.96\\ 
 \hline
 $i=70^{\circ}$ & $42.02\pm 0.02$ & $149.91\pm 0.10$ & 16.56 & $139.34\pm 15.35$ & $9.52\pm 1.33$ & $2.31$ & $89.30\pm13.93$ & $16.26\pm3.18$ & 2.37 & $0.00459\pm0.00482$ & $31.22\pm32.20$ & $0.01\pm0.31$ & 2.48 \\
 \hline
 $i=78^{\circ}$ & $40.02\pm0.02$ & $149.89\pm0.10$ & 14.02 & $101.11\pm6.05$ & $6.17\pm0.55$ & 0.98 & $60.59\pm3.91$ & $9.63\pm0.93$ & 0.97 & $0.00611\pm0.00270$ & $16.54\pm5.67$ & $0.04\pm0.14$ & 1.08 \\
 \hline
 $\rm{PA=293^{\circ}}$ & $40.03\pm0.03$ & $149.81\pm0.19$ & 7.77 & $114.17\pm8.09$ & $7.16\pm0.72$ & 3.07 & $67.74\pm5.24$ & $11.03\pm1.22$ & 3.03 & $0.00050\pm0.00110$ & $150.00\pm439.26$ & $0.54\pm0.13$ & 2.42 \\
 \hline
 $\rm{PA=299^{\circ}}$ & $42.01\pm0.01$ & $149.92\pm0.08$ & 14.73 & $121.89\pm12.03$ & $8.22\pm1.07$ & 2.09 & $74.34\pm9.12$ & $13.16\pm2.13$ & 2.09 & $0.00075\pm0.00145$ & $150.00\pm448.40$ & $0.41\pm0.15$ & 1.97 \\
 \hline
 free center & $42.01\pm0.01$ & $149.93\pm0.07$ & 22.02 & $128.05\pm10.93$ & $8.59\pm0.96$ & 1.72 & $79.36\pm8.35$ & $14.06\pm1.94$ & 1.80 & $0.00501\pm0.00377$ & $26.64\pm18.43$ & $0.00\pm0.23$ & $1.95$ \\
 \hline
 $\gamma^{3.6}_*=0.25$ & $42.01\pm0.01$ & $149.93\pm0.07$ & 17.16 & $102.51\pm6.05$ & $6.09\pm0.55$ & 1.78 & $61.18\pm3.75$ & $9.45\pm0.88$ & 1.72 & $0.00157\pm0.00107$ & $49.18\pm37.18$ & $0.39\pm0.09$ & 1.46 \\
 \hline
 $\gamma^{3.6}_*=1$ & $40.01\pm0.01$ & $149.92\pm0.08$ & 19.46 & $171.23\pm32.53$ & $13.10\pm2.84$ & 3.17 & $135.77\pm48.18$ & $28.50\pm11.31$ & 3.27 & $0.00359\pm0.00489$ & $50.81\pm88.27$ & $0.00\pm0.37$ & 3.36 \\
 \hline
 $\rm{gas}\,z_0=50\,pc$ & $42.01\pm0.01$ & $149.93\pm0.07$ & 17.67 & $124.61\pm11.30$ & $8.34\pm1.00$ & 2.47 & $75.78\pm8.20$ & $13.32\pm1.91$ & 2.47 & $0.00107\pm0.00152$ & $100.00\pm202.98$ & $0.37\pm0.15$ & 2.33 \\
 \hline
 $\rm{gas}\,z_0=500\,pc$ & $42.01\pm0.01$ & $149.93\pm0.07$ & 15.85 & $115.66\pm8.97$ & $7.37\pm0.79$ & 1.73 & $69.27\pm5.94$ & $11.50\pm1.38$ & 1.72 & $0.00186\pm0.00155$ & $50.09\pm47.04$ & $0.31\pm0.14$ & 1.72 \\
 \hline
 $\rm{star}\,z_0=100\,pc$ & $42.01\pm0.01$ & $149.94\pm0.06$ & 19.95 & $116.66\pm9.08$ & $7.62\pm0.81$ & 2.23 & $69.97\pm5.88$ & $11.92\pm1.38$ & 2.22 & $0.00103\pm0.00125$ & $100.00\pm173.08$ & $0.38\pm0.12$ & 2.06 \\
 \hline
 $\rm{star}\,z_0=500\,pc$ & $42.01\pm0.01$ & $149.94\pm0.06$ & 19.44 & $114.79\pm8.57$ & $7.43\pm0.77$ & 1.82 & $68.78\pm5.67$ & $11.61\pm1.33$ & 1.81 & $0.00149\pm0.00131$ & $63.41\pm69.36$ & $0.34\pm0.12$ & 1.76 \\
 \hline
 non-circular & $42.02\pm0.02$ & $149.90\pm0.10$ & 10.30 & $123.21\pm10.08$ & $8.23\pm0.90$ & 0.82 & $75.67\pm7.82$ & $13.31\pm1.82$ & 0.91 & $0.00522\pm0.00655$ & $23.86\pm23.04$ & $0.00\pm0.42$ & 1.06 \\
 \hline
 \hline
\end{tabular}
\end{table}
\end{landscape}


\bsp	
\label{lastpage}
\end{document}